\documentclass[a4paper,fleqn]{cas-dc}

\usepackage[numbers]{natbib}
\usepackage{xcolor}
\usepackage{verbatim}
\usepackage{tcolorbox}
\usepackage{hyperref}
\usepackage{courier}
\usepackage{caption}
\usepackage{subcaption}
\usepackage{amsmath}
\usepackage{booktabs}
\usepackage{pifont}
\hypersetup{
colorlinks=true,
linkcolor=blue,
anchorcolor=blue,
urlcolor=blue,
citecolor=blue,
}

\DeclareMathOperator*{\argmax}{argmax}

\def\tsc#1{\csdef{#1}{\textsc{\lowercase{#1}}\xspace}}
\tsc{WGM}
\tsc{QE}
\tsc{EP}
\tsc{PMS}
\tsc{BEC}
\tsc{DE}

\begin{document}
\let\WriteBookmarks\relax
\def\floatpagepagefraction{1}
\def\textpagefraction{.001}
\shorttitle{\scriptsize{Diverse Title Generation for Stack Overflow Posts with Multiple Sampling Enhanced Transformer}}
\shortauthors{Fengji Zhang et al.}

\title [mode = title]{Diverse Title Generation for Stack Overflow Posts with Multiple Sampling Enhanced Transformer}

\author{Fengji Zhang\textsuperscript{\textit{a}}}[style=chinese]
\ead{zhangfengji@whu.edu.cn}
\address{\textsuperscript{\textit{a}}School of Computer Science, Wuhan University, Wuhan, China}

\author{Jin Liu\textsuperscript{\textit{a}}}[style=chinese]
\ead{jinliu@whu.edu.cn}

\corref{mycorrespondingauthor1}
\cormark[1]

\author{Yao Wan\textsuperscript{\textit{b}}}[style=chinese]
\ead{wanyao@hust.edu.cn}
\address{\textsuperscript{\textit{b}}School of Computer Science and Technology, Huazhong University of Science and Technology, Wuhan, China}

\author{Xiao Yu\textsuperscript{\textit{c}}}[style=chinese]
\ead{xiaoyu@whut.edu.cn}
\address{\textsuperscript{\textit{c}}School of Computer Science and Artificial Intelligence, Wuhan University of Technology, Wuhan, China}

\corref{mycorrespondingauthor1}
\cormark[1]

\author{Xiao Liu\textsuperscript{\textit{d}}}[style=chinese]
\ead{xiao.liu@deakin.edu.au}
\address{\textsuperscript{\textit{d}}School of Information Technology, Deakin University, Geelong, Australia}

\author{Jacky Keung\textsuperscript{\textit{e}}}[style=chinese]
\ead{jacky.keung@cityu.edu.hk}
\address{\textsuperscript{\textit{e}}Department of Computer Science, City University of Hong Kong, Hong Kong, China}

\cortext[cor1]{Corresponding author}

\begin{abstract}
Stack Overflow is one of the most popular programming communities where developers can seek help for their encountered problems. Nevertheless, if inexperienced developers fail to describe their problems clearly, it is hard for them to attract sufficient attention and get the anticipated answers. To address such a problem, we propose M\textsubscript{3}NSCT5, a novel approach to automatically generate multiple post titles from the given code snippets. Developers may take advantage of the generated titles to find closely related posts and complete their problem descriptions.
M\textsubscript{3}NSCT5 employs the CodeT5 backbone, which is a pre-trained Transformer model having an excellent language understanding and generation ability. To alleviate the ambiguity issue that the same code snippets could be aligned with different titles under varying contexts, we propose the maximal marginal multiple nucleus sampling strategy to generate multiple high-quality and diverse title candidates at a time for the developers to choose from.
We build a large-scale dataset with 890,000 question posts covering eight programming languages to validate the effectiveness of M\textsubscript{3}NSCT5. The automatic evaluation results on the BLEU and ROUGE metrics demonstrate the superiority of M\textsubscript{3}NSCT5 over six state-of-the-art baseline models. Moreover, a human evaluation with trustworthy results also demonstrates the great potential of our approach for real-world application. 

\end{abstract}

\begin{keywords}
\sep Stack Overflow
\sep Title Generation
\sep CodeT5
\sep Nucleus Sampling
\sep Maximal Marginal Ranking
\end{keywords}

\maketitle
\section{Introduction}
\label{introduction}

Stack Overflow (SO) is one of the most popular Question\&Answering websites for developers to seek answers to programming problems. However, it remains a challenge~\cite{Chatterjee2020FindingHW, mondal2021early, Rubei2020PostFinderMS} to help developers write high-quality question posts that attract enough attention from potential experts. Especially, non-native English speakers or inexperienced developers may struggle to clearly describe their encountered problems, let alone summarize the problems into informative titles. One way for developers to write better question posts is to first search for related posts with the problematic code snippets and then complete their problem descriptions and post titles. Nonetheless, previous studies~\cite{gao2020generating, zhang2022improving, liu2022sotitle, Gao2022IKW} demonstrated the unsatisfying performance of the commonly used retrieval methods like TF-IDF and BM25~\cite{robertson2009probabilistic} on searching related posts with given code snippets. First, such retrieval methods calculate the lexical overlap and ignore the essential semantic similarity. Second, different from natural language queries, code snippets usually have very long contexts and plentiful user-defined tokens, making it hard to extract lexical features.

\begin{figure}[pos=h]
	\centering
	\includegraphics[scale=.4]{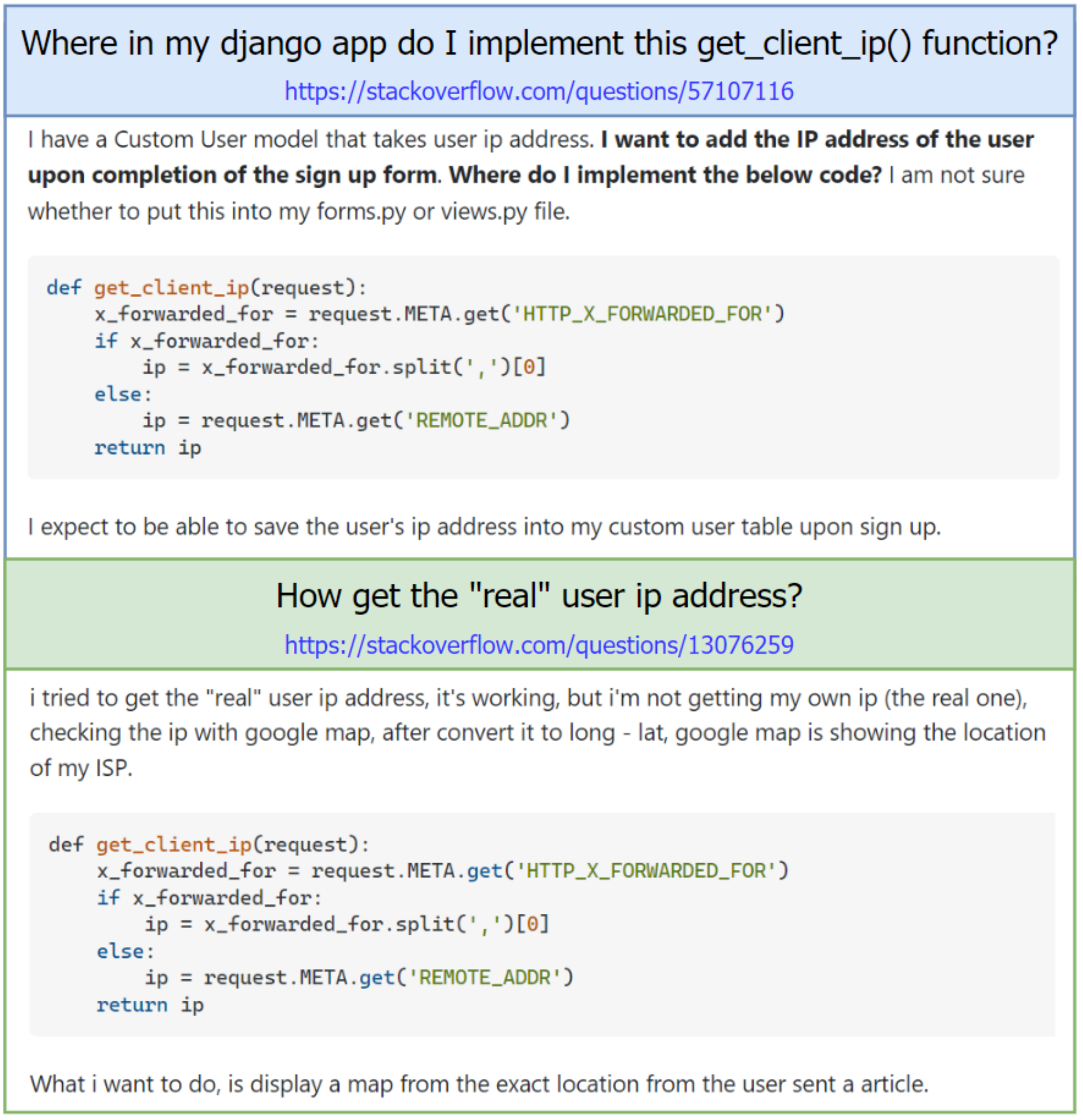}
	\caption{Illustration of the \textit{ambiguity} issue --- posts with the same code snippets have different titles under varying contexts.}
	\label{figure:ambiguity}
\end{figure}

Recently, Gao et al.~\cite{gao2020generating} proposed an end-to-end generation model to automatically produce post titles with the given code snippets. First, they train an LSTM (Long Short Term Memory)~\cite{hochreiter1997long} model on a large-scale dataset collected from Stack Overflow, which contains pairs of code snippets and post titles. Then, a developer could provide the model with code snippets to get a generated post title that summarizes the problem. The generated titles are coherent and informative, which will help developers understand their problems and find related posts more easily. However, as suggested by Liu et al.~\cite{liu2022sotitle}, the same code snippets could be aligned with different titles under varying contexts, which we denote as the \textit{ambiguity} issue. For example, in Figure \ref{figure:ambiguity}, the two SO posts ask different questions and have different titles. However, they have the same code snippets that implement the Python function \textit{get\_client\_ip}. 
Liu et al.~\cite{liu2022sotitle} then proposed to tackle the issue by leveraging the surrounding text descriptions in the post body to eliminate the semantic ambiguity of code snippets. 
Nonetheless, it remains an open challenge to generate the expected post titles when developers cannot provide precise descriptions of their problems.

To mitigate this challenge, we reformulate the title generation task as generating multiple candidate titles simultaneously under the condition that only code snippets are provided. Since code snippets can be ambiguous without the surrounding contexts, we could offer the developers an acceptable amount of candidate titles to choose from. But this will pose a new challenge of improving the diversity of generated titles while keeping the quality so that the titles can nicely summarize the code snippets as well as cover different intentions under varying contexts. 
To this end, we propose M\textsubscript{3}NSCT5, a novel approach to generate high-quality and diverse post titles from the given code snippets. \textbf{M\textsubscript{3}NSCT5} is a hybrid method combining the \textbf{M}aximal \textbf{M}arginal \textbf{M}ultiple \textbf{N}ucleus \textbf{S}ampling strategy and the \textbf{C}ode\textbf{T5} model. 

Specifically, M\textsubscript{3}NSCT5 is based on CodeT5~\cite{Wang2021CodeT5IU}, a Transformer~\cite{Vaswani2017AttentionIA} encoder-decoder model pre-trained on a large-scale code-related corpus, which could better capture the long-range dependencies than LSTM~\cite{khandelwal2018sharp} and generate titles with higher quality. To improve the diversity of generated titles, we apply the nucleus sampling~\cite{holtzman2019curious} method instead of the commonly used beam search during decoding and propose the maximal marginal ranking strategy to  ensure the quality and diversity of the predicted titles. In this way, we can tackle the \textit{ambiguity} issue by offering multiple title candidates for developers to choose from.

To verify the effectiveness of our approach, we conduct the empirical study by raising the following Research Questions (RQs):

\textbf{RQ-1: Does our approach outperform state-of-the-art baselines under automatic evaluation?}
We build a large-scale dataset $D_{so}$ with around 890,000 high-quality SO posts covering eight programming languages. We employ BLEU \cite{Papineni2002BleuAM} and ROUGE \cite{Lin2004ORANGEAM} as the automatic evaluation metrics and choose six baseline models (i.e., BM25~\cite{robertson2009probabilistic}, Code2Que~\cite{gao2020generating}, BART~\cite{lewis2020bart}, CCBERT~\cite{zhang2022improving}, SOTtitle~\cite{liu2022sotitle}, and PLBART~\cite{ahmad2021unified}) for comparison. Experimental results show that M\textsubscript{3}NSCT5 outperforms all the baselines by a large margin, having an around 9\% improvement over the second best performing PLBART baseline on average of different experimental settings. 

\textbf{RQ-2: How effective is our maximal marginal multiple nucleus sampling?}
We compare the performance of our sampling strategy with beam search and vanilla random nucleus sampling. Results show that our method could improve both the quality and diversity of generated titles, especially when the number of output titles is limited to a small value ($\leq 5$), making it suitable for real-world application.

\textbf{RQ-3: What is the performance of our approach under human evaluation?}
To compensate for the non-intuitive automatic evaluation metrics, we recruit six experienced programmers to perform an additional human evaluation. Participants are required to score the titles generated by M\textsubscript{3}NSCT5, PLBART, and BM25 involving three programming languages on the $Readability$, $Correlation$, and $Diversity$ criteria. Results show that our approach also has better performance under human-centered evaluation.

The contributions of this paper are as follows:
\begin{itemize} 
	\item We propose M\textsubscript{3}NSCT5, a novel approach combining the pre-trained CodeT5 model and the maximal marginal multiple nucleus sampling strategy, which could improve the quality and diversity of generated SO titles.
	\item We collect a large-scale dataset containing 890,000 high-quality posts covering eight programming languages and demonstrate the effectiveness of our approach under automatic and human evaluation at different experimental settings.
	\item We have released the source code and processed dataset\footnote{\url{https://github.com/zfj1998/M3NSCT5}}  to facilitate future research.
\end{itemize}

\begin{figure*}
	\centering
	\includegraphics[scale=.47]{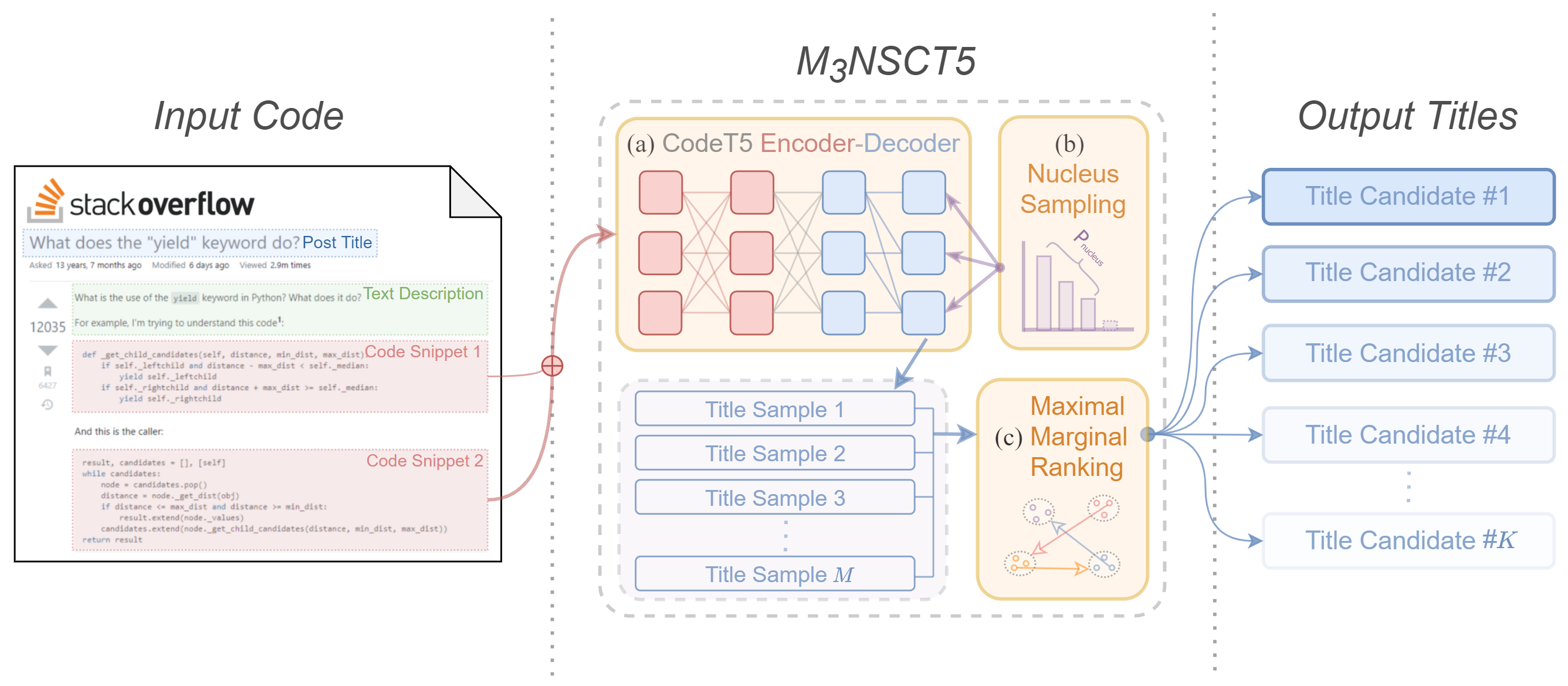}
	\caption{The overall framework of our approach for Stack Overflow post title generation. Given the input code snippets, M\textsubscript{3}NSCT5 can produce multiple title candidates. There are three critical components inside M\textsubscript{3}NSCT5, namely the CodeT5 backbone, the nucleus sampling method, and the maximal marginal ranking strategy.}
	\label{figure:framework}
\end{figure*}

We organize the rest of this paper as follows:
Section \ref{proposed approach} introduces the details of our proposed approach. Section \ref{experimental setup} describes the basic setup of our experiment, including the construction of the experimental dataset, hyper-parameter settings, baseline models, and evaluation metrics. Section \ref{results and analysis} presents the experimental results. Section \ref{related work} introduces the related works. Section \ref{threats to validity} discusses threats to the validity of our work. Finally, we conclude this paper and introduce the future work in Section \ref{conclusion and future work}.

\section{The Proposed Approach}
\label{proposed approach}

Generating post titles from code snippets can be seen as a PL-to-NL (Programming Language to Natural Language) generation task. Figure \ref{figure:framework} illustrates the overall framework of our M\textsubscript{3}NSCT5, a novel end-to-end approach that could improve the quality and diversity of the post titles generated from the code snippets. Specifically, we employ CodeT5 as the backbone, which takes in the code snippets and generates post titles. We further incorporate the nucleus sampling and maximal marginal ranking strategy to produce a set of high-quality and diverse title candidates. The details of our approach are described in this section.

\subsection{CodeT5 Backbone Model}
CodeT5~\cite{Wang2021CodeT5IU} is a state-of-the-art Transformer model pre-trained on a large-scale code-related corpus involving multiple programming languages. It inherits the unified encoder-decoder architecture from T5~\cite{raffel2019exploring}, which has been shown beneficial for generation tasks. 
We follow the pre-train then fine-tune paradigm and further update the trainable parameters $\theta$ of CodeT5 on our task-specific dataset $D_{so}$.

\paragraph{Fine-Tuning:} Our objective is to maximize the probability $P_\theta(Y|X)$ given the input code sequence $X$ and the target title $Y$ from the training dataset.
$X$ and $Y$ are first split into tokens by the default byte-pair encoding tokenizer of CodeT5, then turned into vectors through the embedding layer. Especially, if the input contains multiple code snippets, we concatenate them to a long sequence with the additional \texttt{[NEXT]} identifier. Suppose $X=(x_1...x_{|X|})$ and $Y=(y_1...y_{|Y|})$, where $x_i,y_j\in R^{d_{model}}$. $d_{model}$ is the model hidden size, and $|X| $ and $|Y|$ denote the sequence length with respect to $X$ and $Y$. We feed $X$ to the encoder, which mainly performs bidirectional self-attention to get
\begin{equation}
	C=\mathrm{ENCODER}(X),\label{context_vector}
\end{equation}
where $C=(c_1...c_{|X|})$ and vector $c_i\in R^{d_{model}}$ is the hidden representation of the $i$th input token. 
We then feed the auto-regressive decoder with $C$ and $Y$ to get
\begin{equation}
	G=\mathrm{DECODER}(C, Y),\label{hidden_state}
\end{equation}
where $G=(g_1...g_{|Y|})$ and vector $g_j\in R^{d_{model}}$ represents the hidden state of the $j$th predicted token.
Next, we employ an additional neural layer to map $G$ from the decoder hidden space to the probability distribution over the prediction vocabulary
\begin{equation}
	\mathbf{P}=LinearSoftmax(G),\label{vocabulary_dist}
\end{equation}
where $\mathbf{P}=(P_1...P_{|Y|})$, $P_j\in R^{d_{vocab}}$, $d_{vocab}$ is the vocabulary size, and $LinearSoftmax$ is a linear neural network with the $softmax$ activation function.
Eventually, we can get the loss function for fine-tuning by calculating the average negative log-likelihood
\begin{equation}
	Loss=\frac{1}{|Y|}\sum_{j=1}^{|Y|}-\log{P_j(y_j)},
\end{equation}
where $P_j(y_j)$ is the predicted probability of the $j$th token in the target title.

\paragraph{Inference:} We employ the already fine-tuned model and the auto-regressive decoding method to get the predicted title $\hat{Y}$ token-by-token. To be specific, we first feed the decoder with the start identifier \texttt{<s>} to generate a probability distribution $P_1$ over the vocabulary, which is used for sampling the first predicted token $\hat{y}_1$. After that, we will again take $(\texttt{<s>},\hat{y}_1)$ as the input sequence for the decoder to predict the second token $\hat{y}_2$ by repeating the previous steps. Our model predicts each token in $\hat{Y}$ recursively until encountering the ending identifier \texttt{</s>}.

When generating multiple titles, we follow the parallel manner to save the computation cost. Generally, we first take $M$ start identifiers $(\texttt{<s>}_1...\texttt{<s>}_M)^\intercal$ as the input for decoding. In return, we get the sampled first tokens $(\hat{y}_{1,1}...\hat{y}_{M,1})^\intercal$ for $M$ candidates. Through the auto-regressive decoding method, our model will repeatedly sample tokens at each step until all the candidates meet the ending identifier \texttt{</s>}. Finally, we will get the sampled titles
\begin{equation}
	\hat{Y}_{M,N} = 
	\begin{pmatrix}
		\hat{y}_{1,1} & \hat{y}_{1,2} & \cdots & \hat{y}_{1,N} \\
		\hat{y}_{2,1} & \hat{y}_{2,2} & \cdots & \hat{y}_{2,N} \\
		\vdots  & \vdots  & \ddots & \vdots  \\
		\hat{y}_{M,1} & \hat{y}_{M,2} & \cdots & \hat{y}_{M,N} 
	\end{pmatrix},
\end{equation}
where $\hat{y}_{m,n}$ is the $n$th sampled token of the $m$th candidate title, $N$ is the length of the longest candidate, and all the shorter candidates will be padded to length $N$ with a special \texttt{[PAD]} identifier.

\subsection{Nucleus Sampling}

\begin{figure}
	\centering
	\includegraphics[scale=.23]{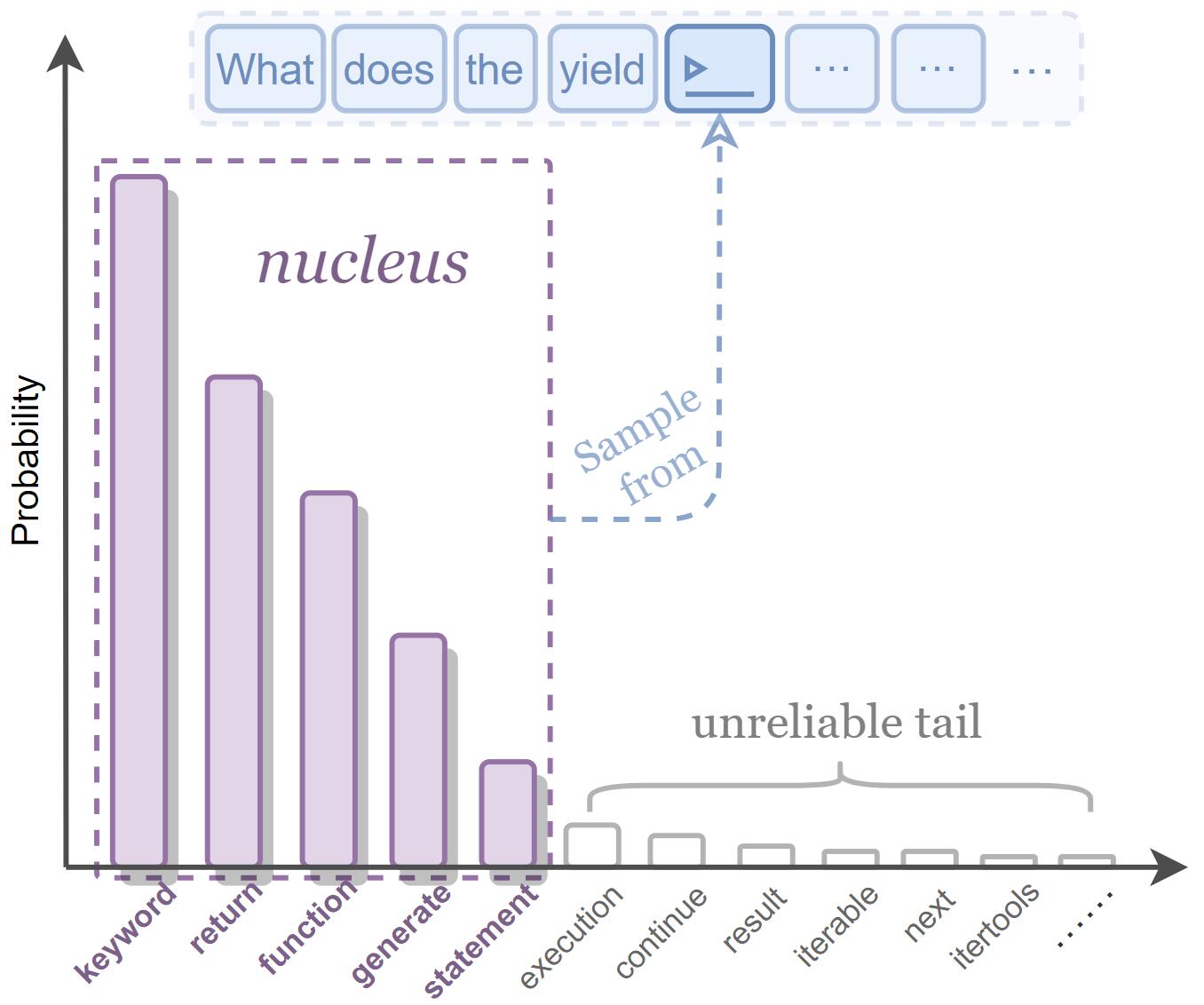}
	\caption{Illustration of applying nucleus sampling to get the next token in the predicted title.}
	\label{figure:nucleus}
\end{figure}

An essential step of the decoding step is to sample the predicted token $\hat{y}$ from its probability distribution $P$ over the vocabulary. The most common sampling method is beam search, whose objective is to maximize the probability $P(\hat{Y})$ over the predicted tokens, where
$P(\hat{Y})=\prod_{r=1}^{|\hat{Y}|}P_r(\hat{y}_r)$.
Nonetheless, the content produced by beam search is found to lack divergence compared with the content written by humans~\cite{holtzman2019curious}. It is because the maximization-based objective always suppresses the occurrence of uncommon phrases.

In this study, we need to ensure the diversity of generated titles so that they can cover different intentions under varying contexts. To this end, we employ the nucleus sampling~\cite{holtzman2019curious} method, whose key intuition is to sample the predicted token from a \textit{nucleus} distribution instead of choosing the token with the highest probability. As shown in Figure \ref{figure:nucleus}, given the already sampled tokens [`What', `does', `the', `yield'], we are now going to choose the next token from the vocabulary distribution. Some tokens in the vocabulary are unlikely to be chosen, such as [`execution', `continue', \dots, `itertools'], which make up the unreliable tail of the distribution. The tokens in the \textit{nucleus}, a minimal subset of the vocabulary that takes up the vast majority of probability mass, are [`keyword', `return', \dots, `statement'], which are most likely to follow the previous token `yield'. Using nucleus sampling, any token in the \textit{nucleus} has the chance to be chosen, which could bring randomness to the sampling process and significantly improve the diversity of generated titles. 

Formally, suppose we are generating the $r$th token $\hat{y}_r$ using nucleus sampling, with the probability distribution $P_r$ over the vocabulary $V$. We first find the minimal \textit{nucleus} set $V^{(p)}\subset V$ such that
\begin{equation}
	\sum_{v\in V^{(p)}}P_r(v)\ge \beta,
\end{equation}
where $v\in V$ and $\beta$ (also denoted as top-$p$) is a hyper-parameter of nucleus sampling ranging from 0.0 to 1.0.
Let $p'=\sum_{v\in V^{(p)}}P_r(v)$. The original distribution $P_r$ can be re-scaled to
\begin{equation}
	{P'_r(v)}=
	\begin{cases}
		P_r(v)/P' &{\rm{if}} \ v\in V^{(p)}\\
		0 &otherwise
	\end{cases},
\end{equation}
and $\hat{y}_r$ will be sampled from the new distribution $P'_r$.

\subsection{Maximal Marginal Ranking}

\begin{figure}
	\centering
	\includegraphics[scale=.32]{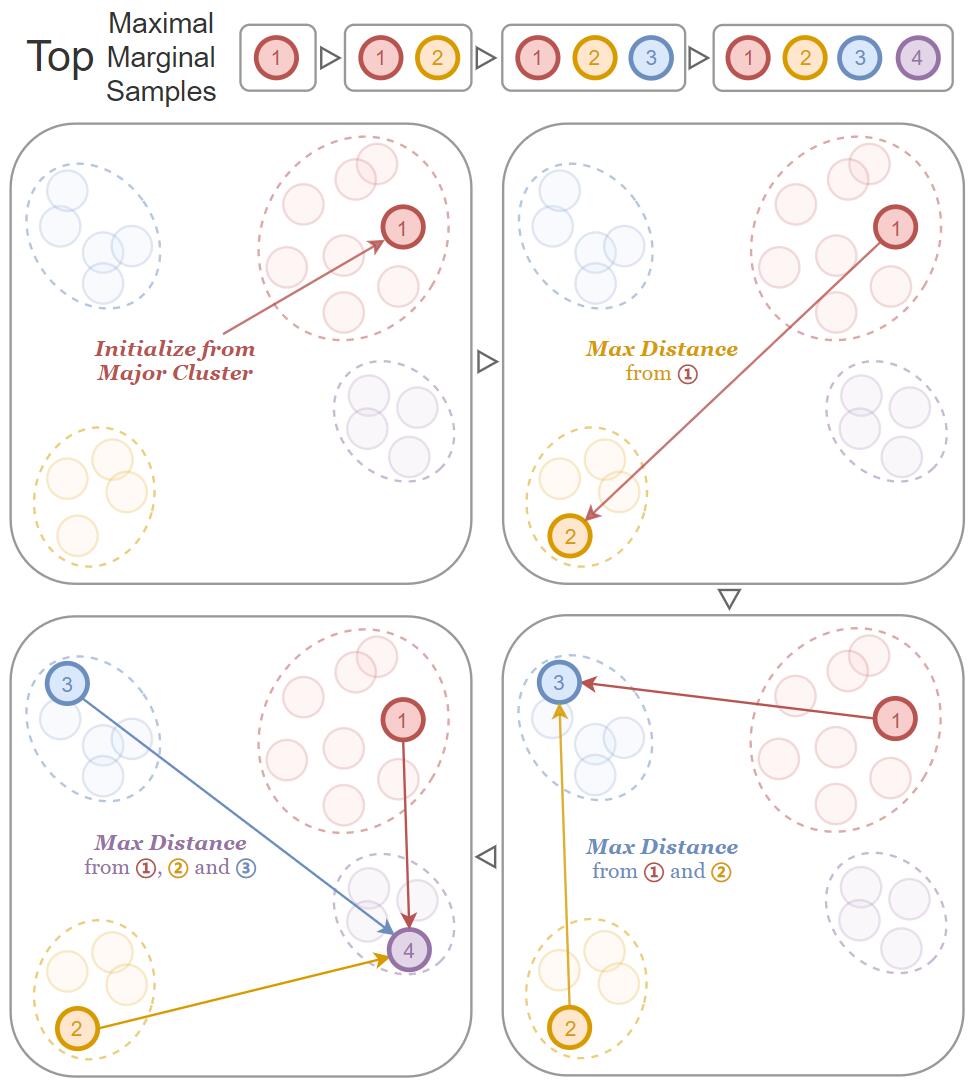}
	\caption{Illustration of the maximal marginal ranking strategy. Nodes marked in four colors denote the title samples grouped into four clusters based on their distance in the space. This figure shows a four-step example of choosing four titles from all the samples: start from an initial title; the following steps are to choose the title that has the maximal distance to the already chosen ones.}
	\label{figure:mmr}
\end{figure}

Nucleus sampling has been successfully applied to the domain of code generation~\cite{li2022competition, chen2021evaluating, fried2022incoder, Hendrycks2021MeasuringCC, Xu2022ASE}. For example, state-of-the-art code generation models AlphaCode~\cite{li2022competition} and OpenAI Codex~\cite{chen2021evaluating} both incorporate nucleus sampling to generate hundreds and thousands of candidate code solutions for each programming problem, which will significantly improve the problem-solving rates. This can be attributed to the randomness brought by the nucleus sampling, which could enlarge the exploration space of pre-trained models and increase the chance of generating high-quality content.
However, due to the random nature of nucleus sampling, there is a high variance in generation quality. A common practice to tackle this issue is to sample multiple times and then choose the best samples~\cite{cobbe2021training, inala2022fault, Shi2022NaturalLT}. For example, AlphaCode~\cite{li2022competition} employs sophisticated filtering and clustering methods over the generated code solutions to narrow the number of candidates so that the target programming problem can be solved within minimum tries.

In this study, we propose a simple yet effective maximal marginal ranking strategy to ensure the diversity and quality of the final predicted titles. We illustrate the rough idea of our ranking strategy in Figure \ref{figure:mmr}, where the nodes in the two-dimensional space represent the title samples produced by nucleus sampling. Furthermore, the nodes (titles) that are similar should have a closer distance. Our goal is to find the top-ranked titles with good diversity and quality from all the samples. First, we need to choose a node to start the ranking process. In the example, we choose the node \ding{172} from the majority cluster of the red color as the initial candidate. Second, we choose the yellow node \ding{173} as another candidate, which has the maximal distance from \ding{172} among all the nodes. Then, we choose the blue node \ding{174} as the next candidate, which has the maximal distance from both \ding{172} and \ding{173}. 
Similarly, we choose the purple node \ding{175} as the fourth candidate, which has the maximal distance from all the previously chosen nodes. In this way, we can include nodes from different clusters to ensure the diversity of chosen titles.
The following introduces the details of our ranking strategy:

\paragraph{Choosing the initial sample:} It is crucial to choose a high-quality initial title to start the ranking process because the maximal marginal ranking objective only guarantees the diversity of chosen titles and is blind to their quality. 
However, discriminating the quality of generated titles is a nontrivial task due to the lack of explicit rules that define `good quality'. To tackle this problem, we adapt the idea of \textit{self-consistency}~\cite{wang2022self} to facilitate selecting the initial title from generated samples. 
The \textit{self-consistency} was proposed to improve the performance of reasoning tasks. Generally, after sampling a set of diverse candidates from the model, the final answer should be the one that is most consistent among the other generated answers.
In this study, we propose to measure the quality of generated titles through the \textit{bi}gram consistency. Precisely, we first extract all the token \textit{bi}grams from the generated titles and then calculate the frequency of each \textit{bi}gram. Finally, we rank the titles based on the average frequency of their \textit{bi}grams, and the top-1 title will be considered the most promising initial sample.

\paragraph{Choosing the next samples:} Suppose we will offer $K$ titles for the developer, our model needs to sample $M$ candidates for ranking, where $M>> K$ (e.g., $M$ is 200 when $K$ is 5). Given the already chosen titles $\mathbf{\hat{Y}}_S\subset \mathbf{\hat{Y}}_M$ ( $\mathbf{\hat{Y}}_S=(\hat{Y}_1...\hat{Y}_S)$, $S<K$), we need to choose the next title $\hat{Y}_{S+1}$ from the rest of candidates $\mathbf{\hat{Y}}_M\setminus \mathbf{\hat{Y}}_S$.
To improve the diversity of chosen titles, we propose to find the one that has the maximal distance from those in $\mathbf{\hat{Y}}_S$,
\begin{equation}
	\hat{Y}_{S+1}=\argmax_{\hat{Y}_m\in \mathbf{\hat{Y}}_M\setminus \mathbf{\hat{Y}}_S}\big(\sum\limits_{\hat{Y}_s\in \mathbf{\hat{Y}}_S}-relevance(\hat{Y}_m, \hat{Y}_s)\big),
\end{equation}
where $relevance(\hat{Y}_m, \hat{Y}_s)$ is computed by the cosine similarity of the bag-of-\textit{bi}gram vectors built from the titles. We repeat this process until the size of $\mathbf{\hat{Y}}_S$ reaches $K$.

\section{Experimental Setup}
\label{experimental setup}

This section introduces the construction of our dataset, the implementation of our model, the baselines for performance comparison, the automatic evaluation metrics, and the criteria for human evaluation.

\subsection{Data Preparation}
\label{data_preparation}
Though previous studies~\cite{gao2020generating,zhang2022improving,liu2022sotitle} have proposed open-sourced datasets for the SO title generation task, there are several drawbacks we still have to overcome. Specifically, Gao et al.~\cite{gao2020generating} only considered the posts with an interrogative title, which account for less than a third of real-world data samples, thus resulting in a biased dataset. While both Zhang et al.~\cite{zhang2022improving} and Liu et al.~\cite{liu2022sotitle} had their published bi-modal posts stripped and tokenized through natural language processing tools, which damaged the lexical and structural information (such as the white spaces and line breaks) of the code snippets. As a result, we re-construct a large-scale dataset $D_{so}$ to perform our experiments.

$D_{so}$ is built on the \textit{SOTorrent} dataset proposed by Baltes et al.~\cite{baltes2018sotorrent}, which is originally used for analyzing the evolution of SO posts. The latest checkpoint of \textit{SOTorrent} contains all the posts from July 2008 to December 2020.
Baltes et al.~\cite{baltes2018sotorrent} extracted the code snippets marked by various notations from post bodies and reserved all the white spaces, line breaks, user-defined identifiers, etc. They also removed the noisy fragments wrongly marked as code in the text blocks. 

Moreover, the previously proposed datasets~\cite{gao2020generating,zhang2022improving,liu2022sotitle} only focus on a few dominant Programming Languages (PLs) with abundant data samples, such as \textit{Python}, \textit{C\#}, \textit{Java}, \textit{JS\small(JavaScript\small)}, and \textit{PHP}.   
In this study, we consider the posts involving eight PLs, including the above popular ones and the minorities (\textit{C}, \textit{Ruby}, and \textit{Go}). Besides, to ensure the quality of training data, we only choose the posts that satisfy the four conditions:

\begin{enumerate}\label{filter condition}
	\item The post is not closed; 
	\item The post has an accepted answer; 
	\item The post gets more than one vote;
	\item The post includes code snippets.
\end{enumerate}

\begin{table}[pos=h]
	\caption{The number of Train/Validation/Test samples in $D_{so}$ with respect to different PLs.}\label{dataset}
	\begin{tabular}{@{\ \ \ }cccc@{\ \ \ }}
		\toprule
		PL & Train & Validation & Test\\ \midrule
		Python & 190,934 & 5,000 & 5,000 \\
		C\# & 175,070 & 5,000 & 5,000 \\
		Java & 162,161 & 5,000 & 5,000 \\
		JS & 151,540 & 5,000 & 5,000 \\
		PHP & 86,729 & 5,000 & 5,000 \\
		C & 29,746 & 3,700 & 3,700 \\
		Ruby & 23,774 & 3,000 & 3,000\\
		Go & 6,820 & 850 & 850\\ \midrule[0.2pt]
		Total & 826,774 & 32,550 & 32,550\\
		\bottomrule
	\end{tabular}
\end{table}

As for data partitioning, we separate the filtered posts in chronological order, where the latest posts are randomly grouped into validation and test sets, and the rest are for training. This is reasonable because our model should take the past data for training and is applied to new questions in the real-world scenario. We set the number of validation and test samples to 5,000 with respect to different PLs. For the languages with insufficient data, we set their proportions of validation and test sets to 10\%.
In the end, we get the large-scale and high-quality dataset $D_{so}$ for the SO title generation task. The statistics of $D_{so}$ is summarized in Table \ref{dataset}.

\subsection{Implementation Details}
We implement M\textsubscript{3}NSCT5 with the transformers\footnote{\url{https://huggingface.co/docs/transformers/index}} library and the pre-trained model checkpoint\footnote{\url{https://huggingface.co/Salesforce/codet5-base}} of CodeT5, which consists of 12 encoder layers and 12 decoder layers with a hidden size of 768. We optimize all the trainable parameters through AdamW~\cite{loshchilov2018decoupled}, with an initial learning rate of $5\times10^{-5}$ scheduled by the linear warm-up. We employ the default byte-pair encoding tokenizer of CodeT5, whose vocabulary size is 32,100. We have two Tesla V100 (16GB memory) GPUs for training, where each one could hold a data batch size of 8. We further increase the overall batch size to 32 by gradient accumulation. The model is set to train for ten epochs, and we employ the early stopping strategy to avoid overfitting. Finally, we set top-$p$ in the nucleus sampling to 0.8, temperature to 1, and the number of sampled candidates to 200 during decoding.

\subsection{Baselines}
\label{baselines}
To demonstrate the effectiveness of our approach, we choose several state-of-the-art baseline methods for comparison. We give a brief introduction to these approaches and their experimental settings.
\begin{enumerate}[(1)]
\item \textbf{BM25}~\cite{robertson2009probabilistic} is a widely used ranking function in information retrieval systems. It could estimate the relevance of documents for a given search query. The basic idea of BM25 is to rank the referencing documents based on the overlapping query terms, thus ignoring their correlation within the document. Our study adopts this method to retrieve the most relevant posts in the training dataset given the testing code snippets. We could select one or more best matches for each query as the predicted title candidates. We take advantage of the ready-to-use Elasticsearch engine\footnote{\url{https://www.elastic.co/elasticsearch/}} to implement this retrieval baseline, whose default similarity ranking algorithm is BM25.

\item \textbf{Code2Que} was proposed by Gao et al.~\cite{gao2020generating} to generate SO titles from given code snippets. It is an end-to-end model with the LSTM~\cite{hochreiter1997long} encoder-decoder architecture. Its encoder is a multi-layer bidirectional LSTM network that sequentially handles the input code tokens, while its decoder is the single-layer LSTM that recursively returns the predicted tokens. Moreover, Code2Que incorporates the copy~\cite{See2017GetTT} mechanism to allow the decoder to focus on more relevant parts of the input and facilitate capturing some rare but important tokens, and the coverage~\cite{Tu2016ModelingCF} mechanism to discourage generating meaningless repetitions. We employ the OpenNMT\footnote{\url{https://opennmt.net}} library to reproduce this baseline method.

\item \textbf{BART}~\cite{lewis2020bart} is a pre-trained Transformer model that achieves state-of-the-art results on a range of NL tasks, especially abstractive summarization, question answering, and machine translation. 
Unlike the previous successful pre-trained language models BERT~\cite{kenton2019bert} (only with the Transformer encoder) and GPT~\cite{radford2018improving} (only with the Transformer decoder), BART employs a standard encoder-decoder architecture and proposes specially designed denoising objectives for pre-training. As a result, BART could improve the performance over previous work when fine-tuned for both text understanding and generation tasks. We reproduce this baseline using its pre-trained model checkpoint\footnote{\url{https://huggingface.co/facebook/bart-base}}.

\item \textbf{CCBERT} was proposed by Zhang et al.~\cite{zhang2022improving}, which is also used for SO title generation but takes bi-modal content (code snippets and text descriptions in the post body) as the model input. CCBERT is a Transformer model equipped with CodeBERT~\cite{feng2020codebert} and an additional copy attention layer. Specifically, CodeBERT is a Transformer encoder pre-trained on a vast scale NL-PL bi-modal corpus, thus having an excellent ability to parse the overall context of SO posts. The copy attention layer is an adapted version of the copy mechanism~\cite{See2017GetTT} for the Transformer architecture, which helps the model focus on input tokens during decoding. Zhang et al.~\cite{zhang2022improving} showed the superiority of CCBERT over Code2Que and BART using their collected dataset.
We take advantage of their published source code and the pre-trained model checkpoint\footnote{\url{https://huggingface.co/microsoft/codebert-base}} of CodeBERT to reproduce this baseline.

\item \textbf{SOTitle} was proposed by Liu et al.~\cite{liu2022sotitle}, which is another novel approach used for SO title generation. The backbone of SOTitle is the pre-trained T5~\cite{raffel2019exploring} model, which follows the Transformer encoder-decoder architecture and employs a transfer learning technique that unifies all text-based language problems into a text-to-text paradigm. T5 was pre-trained on a large-scale corpus crawled from the web and achieved state-of-the-art performance on various NL tasks. Liu et al.~\cite{liu2022sotitle} fine-tuned T5 on their collected SO dataset and reported it could outperform Code2Que and BART. We use their published source code and the pre-trained model checkpoint\footnote{\url{https://huggingface.co/t5-base}} of T5 to reproduce this baseline.

\item \textbf{PLBART}~\cite{ahmad2021unified} is a specialized version of the BART model, whose name is the abbreviation for "Program and Language BART". It also employs the Transformer encoder-decoder architecture and applies denoising objectives for pre-training. PLBART was proposed to produce multilingual representations applicable to NL-PL understanding and generation tasks. It was pre-trained on a large-scale bi-modal corpus collected from GitHub and Stack Overflow, then fine-tuned to downstream applications. Results showed that PLBART could outperform state-of-the-art models in a wide range of tasks, especially code summarization and translation. We reproduce this baseline using its pre-trained model checkpoint\footnote{\url{https://huggingface.co/uclanlp/plbart-base}}.
\end{enumerate}

\subsection{Evaluation Methods}
We believe a high-quality post title should have good readability and a strong correlation with the post body. Manual evaluation is the ideal way to measure these criteria. Nevertheless, considering the tremendous scale of our dataset, it is necessary to perform an automatic evaluation. An additional human evaluation is performed on a small subset of test samples to demonstrate the intuitive quality of titles generated by our model.

\subsubsection{Automatic Evaluation}
Following the previous studies~\cite{gao2020generating, zhang2022improving, liu2022sotitle}, we automatically evaluate the quality of title generation by measuring the similarity between the generated titles and the original titles paired with the input code snippets. We employ two kinds of text similarity measuring methods, namely BLEU~\cite{Papineni2002BleuAM} and ROUGE~\cite{Lin2004ROUGEAP}.

BLEU originates from machine translation tasks, which mainly calculates the lexical overlap between sentences through n-gram precision. It also incorporates the \textit{brevity penalty} to penalize the behavior of generating short sentences for higher precision scores. In our experiments, we use the BLEU-4 score calculated with 1/2/3/4-gram. Besides, we apply a smoothing method introduced by Lin et al.~\cite{Lin2004ORANGEAM} to prevent negative scores caused by excessive short sentences. We denote the smoothed method as BLEUS-4 and take advantage of the NLTK\footnote{\url{http://www.nltk.org/_modules/nltk/translate/bleu_score.html}} library for implementation.

ROUGE is a set of metrics commonly used in text summarization, mainly focusing on the n-gram recall. In our experiments, we employ three ROUGE-family metrics, including the ROUGE-1/2 scores that are calculated with 1/2-gram co-occurrence and the ROUGE-L score that concerns the longest common subsequence. Moreover, we take advantage of an open source library\footnote{\url{https://pypi.org/project/rouge}} for implementation.

\subsubsection{Human Evaluation}
\label{human_evaluation}
In practice, a high-quality post title can be written in different styles. It is hard to tell the actual quality of a generated title based on its similarity with a single reference. Therefore, we perform an additional evaluation on three human-centered criteria. As described in Table \ref{human_eval_criteria}, each criterion can be quantified by a score number. Specifically, \textit{Readability} measures the grammaticality and fluency of a title, while \textit{Correlation} considers the consistency between a title and its corresponding post body. \textit{Diversity} is the number of titles that have distinct meanings.

We recruit six students for human evaluation, asking them to review the titles generated by M\textsubscript{3}NSCT5, PLBART, and BM25. Specifically, all the participants are experienced programmers familiar with Stack Overflow. We assign each participant 100 random post samples where each post is paired with nine titles generated by the three approaches (i.e., the output number $K$=$3$). The participants are evenly divided into three groups according to their preferred programming languages (including the popular \textit{Python} and \textit{Java} languages, as well as the low-resource \textit{Go} language).
During the evaluation, participants do not know the titles are generated by which approach, and they should tell the \textit{Diversity}, \textit{Readability}, and \textit{Correlation} scores of each sample according to the scoring standard in Table \ref{human_eval_criteria}. Then, we take the average score of each two participants in the same group and report the results by different PLs. Finally, we employ Cohen's Kappa~\cite{cohen1960coefficient} to measure the agreement between the two participants in each group.

\begin{table}[pos=h]
	\centering
	\caption{The criteria used for human evaluation.}
	\label{human_eval_criteria}
	\begin{tabular}{@{\ \ }cl@{\ }}
		\toprule
		Criteria  &  Scoring Standard \\ \midrule[0.6pt]
		Readability & 
		\begin{tabular}[c]{@{}l@{}}\textbf{1} $\Rightarrow$ Very hard to read and understand \\\textbf{2} $\Rightarrow$ Just readable and understandable \\\textbf{3} $\Rightarrow$ Very easy to read and understand\end{tabular} \\ \midrule[0.4pt]
		Correlation & 
		\begin{tabular}[c]{@{}l@{}}\textbf{1} $\Rightarrow$ Totally digress from the key points \\\textbf{2} $\Rightarrow$ Relevant to the key points \\\textbf{3} $\Rightarrow$ Exactly match the key points\end{tabular} \\ \midrule[0.4pt]
		Diversity & 
		\begin{tabular}[c]{@{}l@{}}\textbf{[1,K]} $\Rightarrow$ The number of non-redundant titles\end{tabular} \\ \bottomrule
	\end{tabular} 
\end{table}

\subsubsection{Evaluation on Multiple Outputs}
\label{evaluation_on_multiple}
Finally, we introduce the evaluation method when the model outputs multiple titles for a single input. Suppose the output number is $K$. We first calculate the scores of all the titles on a specific \textit{Metric} and then take the highest score as the result, which is denoted as \textit{Metric}@$K$. In this way, we can get the BLEU$K$, ROUGE@$K$, Readability@$K$, Correlation@$K$, and Diversity@$K$ that are used for our experiments.

\subsection{Research Questions}

\begin{table*}
	\centering
	\caption{The automatic evaluation results of M\textsubscript{3}NSCT5 and six baselines on the test dataset with respect to different PLs. The values in the table are the average scores, and $K$ is the number of output titles. B4, R1, R2, and RL are the abbreviations of BLEUS-4, ROUGE-1, ROUGE-2, and ROUGE-L.}
	\label{baselines_k_equal_1}
	\begin{subtable}[t]{0.495\linewidth}
	    \centering
		\caption{Python}
		\begin{tabular}{@{\ \ }cccccc@{\ }}
			\toprule
			Setting & Model & B4@$K$ & R1@$K$ & R2@$K$ & RL@$K$ \\
			\midrule
			\multirow{7}{*}{$K$=1}
			& \scriptsize BM25 & 6.95 & 11.42 & 1.53 & 10.70 \\
			& \scriptsize Code2Que & 12.06 & 23.07 & 6.61 & 22.17 \\
			& \scriptsize BART & 12.76 & 24.98 & 7.56 & 23.13 \\
			& \scriptsize CCBERT & 12.98 & 25.66 & 8.12 & 24.15 \\
			& \scriptsize SOTitle & 12.90 & 25.45 & 7.85 & 23.63 \\
			& \scriptsize PLBART & 13.05 & 26.55 & 8.50 & 24.69 \\
			& \scriptsize \textbf{M\textsubscript{3}NSCT5} & \textbf{13.34} & \textbf{28.65} & \textbf{9.68} & \textbf{26.44} \\
			\specialrule{\heavyrulewidth}{\aboverulesep}{2\belowrulesep}
		\end{tabular}
	\end{subtable}
	\begin{subtable}[t]{0.495\linewidth}
	    \centering
		\caption{C\#}
		\begin{tabular}{@{\ \ }cccccc@{\ }}
			\toprule
			Setting & Model & B4@$K$ & R1@$K$ & R2@$K$ & RL@$K$ \\
			\midrule
			\multirow{7}{*}{$K$=1}
			& \scriptsize BM25 & 6.36 & 9.93 & 1.85 & 9.53 \\
			& \scriptsize Code2Que & 9.91 & 17.45 & 5.05 & 17.55 \\
			& \scriptsize BART & 11.48 & 20.95 & 6.81 & 19.99 \\
			& \scriptsize CCBERT & 11.05 & 20.30 & 6.79 & 19.62 \\
			& \scriptsize SOTitle & 11.52 & 20.91 & 6.67 & 19.98 \\
			& \scriptsize PLBART & 11.61 & 22.58 & 7.73 & 21.68 \\
			& \scriptsize \textbf{M\textsubscript{3}NSCT5} & \textbf{12.16} & \textbf{25.06} & \textbf{9.10} & \textbf{23.75} \\
			\specialrule{\heavyrulewidth}{\aboverulesep}{2\belowrulesep}
		\end{tabular}
	\end{subtable}
	\begin{subtable}[t]{0.495\linewidth}
	    \centering
		\caption{Java}
		\begin{tabular}{@{\ \ }cccccc@{\ }}
			\toprule
			Setting & Model & B4@$K$ & R1@$K$ & R2@$K$ & RL@$K$ \\
			\midrule
			\multirow{7}{*}{$K$=1}
			& \scriptsize BM25 & 6.43 & 10.68 & 1.49 & 10.14 \\
			& \scriptsize Code2Que & 10.51 & 19.49 & 5.24 & 19.25 \\
			& \scriptsize BART & 11.53 & 22.32 & 6.48 & 21.11 \\
			& \scriptsize CCBERT & 11.46 & 22.13 & 6.89 & 21.23 \\
			& \scriptsize SOTitle & 11.63 & 22.55 & 6.58 & 21.36 \\
			& \scriptsize PLBART & 11.72 & 24.14 & 7.56 & 22.94 \\
			& \scriptsize \textbf{M\textsubscript{3}NSCT5} & \textbf{12.37} & \textbf{26.07} & \textbf{8.60} & \textbf{24.46} \\
			\specialrule{\heavyrulewidth}{\aboverulesep}{2\belowrulesep}
		\end{tabular}
	\end{subtable}
	\begin{subtable}[t]{0.495\linewidth}
	    \centering
		\caption{JavaScript}
		\begin{tabular}{@{\ \ }cccccc@{\ }}
			\toprule
			Setting & Model & B4@$K$ & R1@$K$ & R2@$K$ & RL@$K$ \\
			\midrule
			\multirow{7}{*}{$K$=1}
			& \scriptsize BM25 & 6.60 & 10.57 & 1.43 & 10.03 \\
			& \scriptsize Code2Que & 11.29 & 20.83 & 5.78 & 20.44 \\
			& \scriptsize BART & 12.21 & 23.45 & 6.68 & 22.15 \\
			& \scriptsize CCBERT & 12.36 & 23.84 & 7.21 & 22.63 \\
			& \scriptsize SOTitle & 12.34 & 23.73 & 6.89 & 22.48 \\
			& \scriptsize PLBART & 12.50 & 24.88 & 7.58 & 23.65 \\
			& \scriptsize \textbf{M\textsubscript{3}NSCT5} & \textbf{12.74} & \textbf{26.96} & \textbf{8.53} & \textbf{25.25} \\
			\specialrule{\heavyrulewidth}{\aboverulesep}{2\belowrulesep}
		\end{tabular}
	\end{subtable}
	\begin{subtable}[t]{0.495\linewidth}
	    \centering
		\caption{PHP}
		\begin{tabular}{@{\ \ }cccccc@{\ }}
			\toprule
			Setting & Model & B4@$K$ & R1@$K$ & R2@$K$ & RL@$K$ \\
			\midrule
			\multirow{7}{*}{$K$=1}
			& \scriptsize BM25 & 7.72 & 12.15 & 1.53 & 11.27 \\
			& \scriptsize Code2Que & 11.14 & 19.97 & 5.14 & 19.28 \\
			& \scriptsize BART & 12.24 & 22.94 & 5.75 & 21.29 \\
			& \scriptsize CCBERT & 12.46 & 23.04 & 6.26 & 21.50 \\
			& \scriptsize SOTitle & 12.40 & 22.87 & 5.76 & 21.14 \\
			& \scriptsize PLBART & 12.48 & 24.06 & 6.53 & 22.45 \\
			& \scriptsize \textbf{M\textsubscript{3}NSCT5} & \textbf{12.91} & \textbf{25.86} & \textbf{7.45} & \textbf{23.82} \\
			\specialrule{\heavyrulewidth}{\aboverulesep}{2\belowrulesep}
		\end{tabular}
	\end{subtable}
	\begin{subtable}[t]{0.495\linewidth}
	    \centering
		\caption{C}
		\begin{tabular}{@{\ \ }cccccc@{\ }}
			\toprule
			Setting & Model & B4@$K$ & R1@$K$ & R2@$K$ & RL@$K$ \\
			\midrule
			\multirow{7}{*}{$K$=1}
			& \scriptsize BM25 & 6.32 & 10.07 & 1.42 & 9.62 \\
			& \scriptsize Code2Que & 9.24 & 16.52 & 4.22 & 16.40 \\
			& \scriptsize BART & 10.68 & 19.83 & 5.62 & 18.97 \\
			& \scriptsize CCBERT & 10.75 & 20.15 & 5.84 & 19.36 \\
			& \scriptsize SOTitle & 10.91 & 20.30 & 5.69 & 19.42 \\
			& \scriptsize PLBART & 10.98 & 21.67 & 6.48 & 20.73 \\
			& \scriptsize \textbf{M\textsubscript{3}NSCT5} & \textbf{11.49} & \textbf{24.18} & \textbf{7.68} & \textbf{22.85} \\
			\specialrule{\heavyrulewidth}{\aboverulesep}{2\belowrulesep}
		\end{tabular}
	\end{subtable}
	\begin{subtable}[t]{0.495\linewidth}
	    \centering
		\caption{Ruby}
		\begin{tabular}{@{\ \ }cccccc@{\ }}
			\toprule
			Setting & Model & B4@$K$ & R1@$K$ & R2@$K$ & RL@$K$ \\
			\midrule
			\multirow{7}{*}{$K$=1}
			& \scriptsize BM25 & 6.87 & 10.98 & 1.44 & 10.31 \\
			& \scriptsize Code2Que & 11.24 & 20.34 & 5.72 & 19.73 \\
			& \scriptsize BART & 12.74 & 23.60 & 7.04 & 22.08 \\
			& \scriptsize CCBERT & 12.80 & 24.36 & 8.00 & 23.12 \\
			& \scriptsize SOTitle & 12.60 & 23.59 & 7.13 & 22.11 \\
			& \scriptsize PLBART & 12.92 & 24.41 & 7.59 & 22.92 \\
			& \scriptsize \textbf{M\textsubscript{3}NSCT5} & \textbf{13.08} & \textbf{26.77} & \textbf{9.31} & \textbf{25.13} \\
			\specialrule{\heavyrulewidth}{\aboverulesep}{2\belowrulesep}
		\end{tabular}
	\end{subtable}
	\begin{subtable}[t]{0.495\linewidth}
	    \centering
		\caption{Go}
		\begin{tabular}{@{\ \ }cccccc@{\ }}
			\toprule
			Setting & Model & B4@$K$ & R1@$K$ & R2@$K$ & RL@$K$ \\
			\midrule
			\multirow{7}{*}{$K$=1}
			& \scriptsize BM25 & 6.74 & 10.56 & 1.40 & 9.87 \\
			& \scriptsize Code2Que & 10.66 & 18.84 & 4.76 & 19.00 \\
			& \scriptsize BART & 12.45 & 22.63 & 6.44 & 21.52 \\
			& \scriptsize CCBERT & 12.54 & 22.61 & 7.22 & 21.76 \\
			& \scriptsize SOTitle & 12.66 & 22.70 & 6.44 & 21.36 \\
			& \scriptsize PLBART & 12.82 & 23.78 & 7.44 & 22.84 \\
			& \scriptsize \textbf{M\textsubscript{3}NSCT5} & \textbf{13.21} & \textbf{25.57} & \textbf{8.85} & \textbf{24.50} \\
			\specialrule{\heavyrulewidth}{\aboverulesep}{2\belowrulesep}
		\end{tabular}
	\end{subtable}
\end{table*}

\begin{table*}
	\centering
	\caption{The automatic evaluation results of M\textsubscript{3}NSCT5, PLBART, and BM25 on the test dataset when $K>1$, where $K$ is the number of output titles. The values in the table are the average scores of the best title among the $K$ title candidates. B4, R1, R2, and RL are the abbreviations of BLEUS-4, ROUGE-1, ROUGE-2, and ROUGE-L.}
	\label{baselines_k_equal_35}
	\begin{subtable}[t]{0.495\linewidth}
	    \centering
		\caption{Python}
		\begin{tabular}{@{\ \ }cccccc@{\ }}
			\toprule
			Setting & Model & B4@$K$ & R1@$K$ & R2@$K$ & RL@$K$ \\
			\midrule
			\multirow{3}{*}{$K$=3}
			& \scriptsize BM25 & 11.18 & 19.26 & 3.56 & 17.95 \\
			& \scriptsize PLBART & 15.11 & 30.81 & 10.76 & 28.67 \\
			& \scriptsize \textbf{M\textsubscript{3}NSCT5} & \textbf{15.94} & \textbf{33.42} & \textbf{11.93} & \textbf{30.96} \\ \midrule 
			\multirow{3}{*}{$K$=5}
			& \scriptsize BM25 & 12.72 & 22.55 & 4.87 & 21.00 \\
			& \scriptsize PLBART & 16.17 & 33.09 & 12.06 & 30.86 \\
			& \scriptsize \textbf{M\textsubscript{3}NSCT5} & \textbf{17.08} & \textbf{35.58} & \textbf{13.28} & \textbf{33.05} \\
			\specialrule{\heavyrulewidth}{\aboverulesep}{2\belowrulesep}
		\end{tabular}
	\end{subtable}
	\begin{subtable}[t]{0.495\linewidth}
	    \centering
		\caption{C\#}
		\begin{tabular}{@{\ \ }cccccc@{\ }}
			\toprule
			Setting & Model & B4@$K$ & R1@$K$ & R2@$K$ & RL@$K$ \\
			\midrule
			\multirow{3}{*}{$K$=3}
			& \scriptsize BM25 & 10.81 & 17.37 & 4.07 & 16.66 \\
			& \scriptsize PLBART & 14.44 & 27.02 & 9.89 & 25.84 \\
			& \scriptsize \textbf{M\textsubscript{3}NSCT5} & \textbf{14.87} & \textbf{29.54} & \textbf{10.87} & \textbf{27.97} \\ \midrule
			\multirow{3}{*}{$K$=5}
			& \scriptsize BM25 & 12.52 & 20.85 & 5.50 & 19.92 \\
			& \scriptsize PLBART & 15.57 & 29.29 & 11.26 & 28.00 \\
			& \scriptsize \textbf{M\textsubscript{3}NSCT5} & \textbf{16.06} & \textbf{31.75} & \textbf{12.14} & \textbf{30.09} \\
			\specialrule{\heavyrulewidth}{\aboverulesep}{2\belowrulesep}
		\end{tabular}
	\end{subtable}
	\begin{subtable}[t]{0.495\linewidth}
	    \centering
		\caption{Java}
		\begin{tabular}{@{\ \ }cccccc@{\ }}
			\toprule
			Setting & Model & B4@$K$ & R1@$K$ & R2@$K$ & RL@$K$ \\
			\midrule
			\multirow{3}{*}{$K$=3}
			& \scriptsize BM25 & 10.67 & 18.12 & 3.24 & 17.14 \\
			& \scriptsize PLBART & 14.47 & 28.25 & 9.42 & 26.73 \\
			& \scriptsize \textbf{M\textsubscript{3}NSCT5} & \textbf{14.99} & \textbf{30.71} & \textbf{10.49} & \textbf{28.89} \\ \midrule
			\multirow{3}{*}{$K$=5}
			& \scriptsize BM25 & 12.41 & 21.60 & 4.56 & 20.46 \\
			& \scriptsize PLBART & 15.47 & 30.45 & 10.61 & 28.86 \\
			& \scriptsize \textbf{M\textsubscript{3}NSCT5} & \textbf{16.21} & \textbf{32.94} & \textbf{11.80} & \textbf{30.99} \\
			\specialrule{\heavyrulewidth}{\aboverulesep}{2\belowrulesep}
		\end{tabular}
	\end{subtable}
	\begin{subtable}[t]{0.495\linewidth}
	    \centering
		\caption{JavaScript}
		\begin{tabular}{@{\ \ }cccccc@{\ }}
			\toprule
			Setting & Model & B4@$K$ & R1@$K$ & R2@$K$ & RL@$K$ \\
			\midrule
			\multirow{3}{*}{$K$=3}
			& \scriptsize BM25 & 10.87 & 18.02 & 3.33 & 17.02 \\
			& \scriptsize PLBART & 14.69 & 29.44 & 9.68 & 27.74 \\
			& \scriptsize \textbf{M\textsubscript{3}NSCT5} & \textbf{15.15} & \textbf{31.46} & \textbf{10.46} & \textbf{29.53} \\ \midrule
			\multirow{3}{*}{$K$=5}
			& \scriptsize BM25 & 12.49 & 21.19 & 4.50 & 20.01 \\
			& \scriptsize PLBART & 15.80 & 31.68 & 11.01 & 29.84 \\
			& \scriptsize \textbf{M\textsubscript{3}NSCT5} & \textbf{16.25} & \textbf{33.64} & \textbf{11.72} & \textbf{31.65} \\
			\specialrule{\heavyrulewidth}{\aboverulesep}{2\belowrulesep}
		\end{tabular}
	\end{subtable}
	\begin{subtable}[t]{0.495\linewidth}
	    \centering
		\caption{PHP}
		\begin{tabular}{@{\ \ }cccccc@{\ }}
			\toprule
			Setting & Model & B4@$K$ & R1@$K$ & R2@$K$ & RL@$K$ \\
			\midrule
			\multirow{3}{*}{$K$=3}
			& \scriptsize BM25 & 12.08 & 19.93 & 3.46 & 18.38 \\
			& \scriptsize PLBART & 14.67 & 28.99 & 8.61 & 26.96 \\
			& \scriptsize \textbf{M\textsubscript{3}NSCT5} & \textbf{15.66} & \textbf{31.75} & \textbf{9.81} & \textbf{29.32} \\ \midrule
			\multirow{3}{*}{$K$=5}
			& \scriptsize BM25 & 13.75 & 23.39 & 4.93 & 21.60 \\
			& \scriptsize PLBART & 15.76 & 31.16 & 9.75 & 29.04 \\
			& \scriptsize \textbf{M\textsubscript{3}NSCT5} & \textbf{16.97} & \textbf{34.53} & \textbf{11.46} & \textbf{31.84} \\
			\specialrule{\heavyrulewidth}{\aboverulesep}{2\belowrulesep}
		\end{tabular}
	\end{subtable}
	\begin{subtable}[t]{0.495\linewidth}
	    \centering
		\caption{C}
		\begin{tabular}{@{\ \ }cccccc@{\ }}
			\toprule
			Setting & Model & B4@$K$ & R1@$K$ & R2@$K$ & RL@$K$ \\
			\midrule
			\multirow{3}{*}{$K$=3}
			& \scriptsize BM25 & 10.72 & 17.61 & 3.28 & 16.76 \\
			& \scriptsize PLBART & 13.62 & 26.33 & 8.49 & 25.09 \\
			& \scriptsize \textbf{M\textsubscript{3}NSCT5} & \textbf{14.09} & \textbf{28.79} & \textbf{9.48} & \textbf{27.21} \\ \midrule
			\multirow{3}{*}{$K$=5}
			& \scriptsize BM25 & 12.49 & 20.96 & 4.49 & 19.83 \\
			& \scriptsize PLBART & 14.78 & 28.61 & 9.73 & 27.25 \\
			& \scriptsize \textbf{M\textsubscript{3}NSCT5} & \textbf{15.40} & \textbf{31.22} & \textbf{10.73} & \textbf{29.46} \\
			\specialrule{\heavyrulewidth}{\aboverulesep}{2\belowrulesep}
		\end{tabular}
	\end{subtable}
	\begin{subtable}[t]{0.495\linewidth}
	    \centering
		\caption{Ruby}
		\begin{tabular}{@{\ \ }cccccc@{\ }}
			\toprule
			Setting & Model & B4@$K$ & R1@$K$ & R2@$K$ & RL@$K$ \\
			\midrule
			\multirow{3}{*}{$K$=3}
			& \scriptsize BM25 & 11.04 & 18.28 & 3.40 & 17.18 \\
			& \scriptsize PLBART & 15.24 & 29.63 & 10.34 & 27.86 \\
			& \scriptsize \textbf{M\textsubscript{3}NSCT5} & \textbf{16.17} & \textbf{32.16} & \textbf{11.49} & \textbf{30.25} \\ \midrule
			\multirow{3}{*}{$K$=5}
			& \scriptsize BM25 & 12.84 & 21.82 & 4.81 & 20.51 \\
			& \scriptsize PLBART & 16.98 & 31.90 & 11.84 & 30.06 \\
			& \scriptsize \textbf{M\textsubscript{3}NSCT5} & \textbf{17.63} & \textbf{34.77} & \textbf{13.14} & \textbf{32.70} \\
			\specialrule{\heavyrulewidth}{\aboverulesep}{2\belowrulesep}
		\end{tabular}
	\end{subtable}
	\begin{subtable}[t]{0.495\linewidth}
	    \centering
		\caption{Go}
		\begin{tabular}{@{\ \ }cccccc@{\ }}
			\toprule
			Setting & Model & B4@$K$ & R1@$K$ & R2@$K$ & RL@$K$ \\
			\midrule
			\multirow{3}{*}{$K$=3}
			& \scriptsize BM25 & 11.56 & 18.30 & 3.42 & 17.30 \\
			& \scriptsize PLBART & 15.26 & 28.58 & 9.19 & 27.13 \\
			& \scriptsize \textbf{M\textsubscript{3}NSCT5} & \textbf{16.22} & \textbf{30.80} & \textbf{10.97} & \textbf{29.39} \\ \midrule
			\multirow{3}{*}{$K$=5}
			& \scriptsize BM25 & 13.11 & 21.58 & 4.71 & 20.33 \\
			& \scriptsize PLBART & 16.48 & 30.64 & 10.23 & 29.12 \\
			& \scriptsize \textbf{M\textsubscript{3}NSCT5} & \textbf{17.69} & \textbf{33.60} & \textbf{12.23} & \textbf{31.74} \\
			\specialrule{\heavyrulewidth}{\aboverulesep}{2\belowrulesep}
		\end{tabular}
	\end{subtable}
\end{table*}

We demonstrate the effectiveness of our model by conducting experiments to answer the following Research Questions (RQs):
\begin{enumerate}[RQ-1]
	\item \textbf{Does our approach outperform state-of-the-art baselines under automatic evaluation?}
	
	\noindent \textbf{Motivation: }In section \ref{baselines}, we have introduced several state-of-the-art models proposed for the SO title generation task (i.e., Code2Que, CCBERT, and SOTitle) as well as the promising approaches for this task (i.e., BM25, BART, and PLBART). 
	This research question explores whether our model could improve the quality of generated titles compared with the existing methods. 
	
	\item \textbf{How effective is our maximal marginal multiple nucleus sampling?}
	
	\noindent \textbf{Motivation: }Apart from applying CodeT5 as our backbone, the novelty of M\textsubscript{3}NSCT5 mainly lies in our elaborate sampling strategy. We use the nucleus sampling instead of beam search and propose the maximal marginal ranking for further performance improvement. This research question aims to investigate the effectiveness of our sampling strategy.
	
	\item \textbf{What is the performance of our approach under human evaluation?}
	
	\noindent \textbf{Motivation: }Automatic metrics mainly evaluate the similarity between the generated titles and the given references. Nevertheless, such similarity does not necessarily correlate to human perceptible quality. This research question aims to demonstrate the intuitive quality of generated titles through human evaluation.
	
\end{enumerate}

\section{Results and Analysis}
\label{results and analysis}

\begin{figure*}
	\centering
	\begin{subfigure}[t]{0.49\linewidth}
		\centering
		\includegraphics[width=\linewidth]{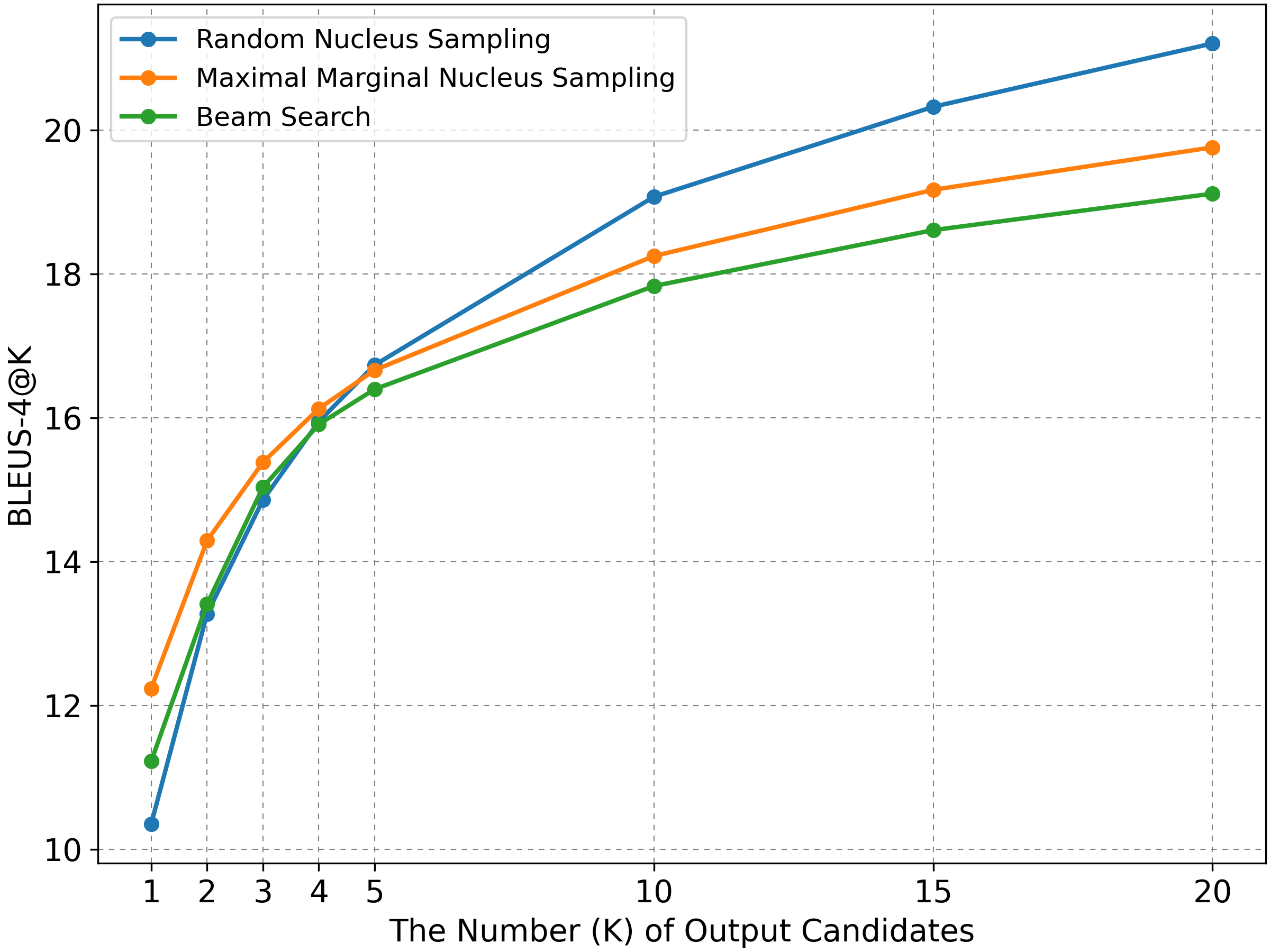}
		\caption{BLEUS-4@$K$}
		\label{bleu}
	\end{subfigure}
	\hfill
	\begin{subfigure}[t]{0.49\linewidth}
		\centering
		\includegraphics[width=\linewidth]{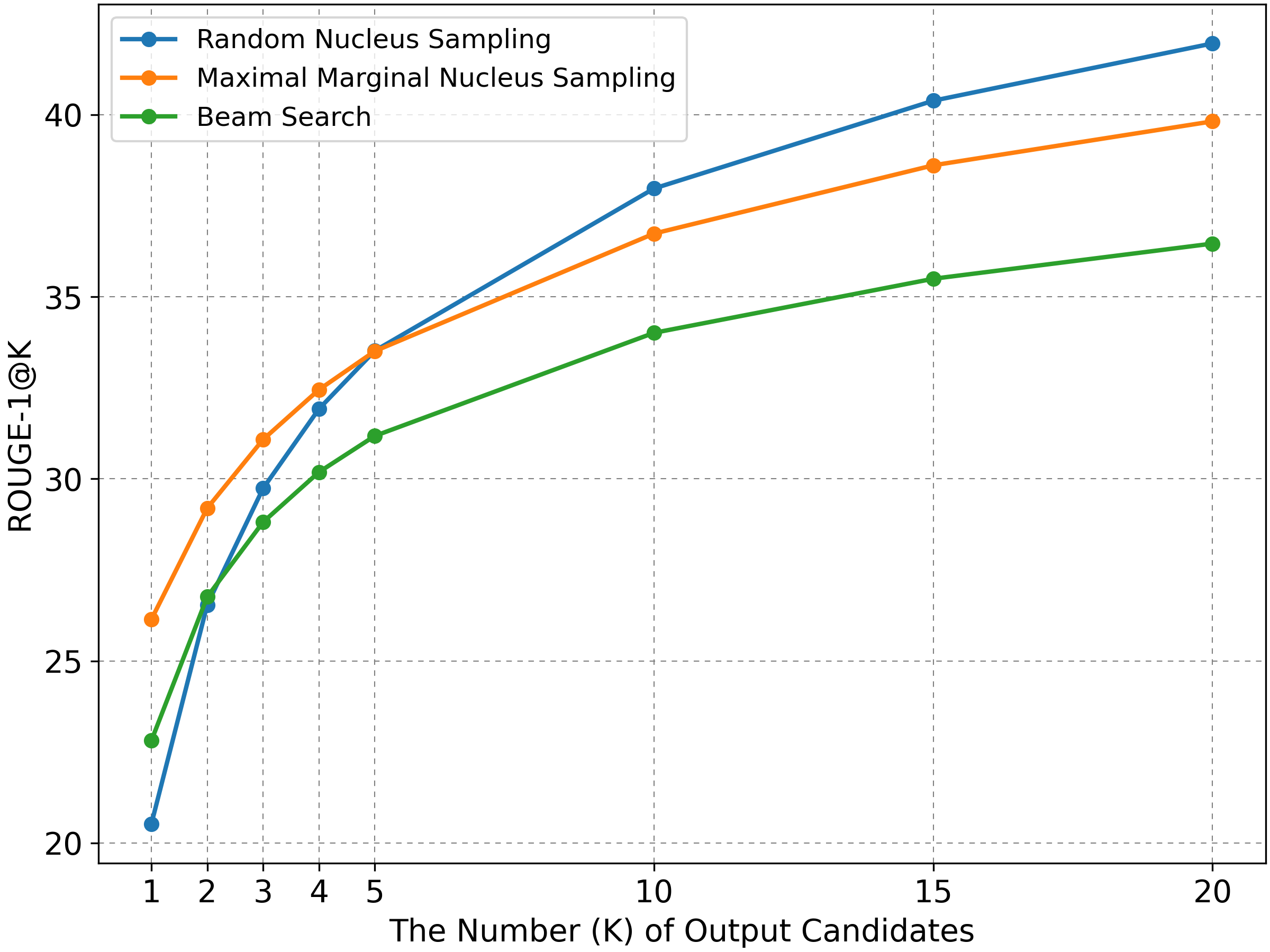}
		\caption{ROUGE-1@$K$}
		\label{rouge1}
	\end{subfigure}
	\hfill
	\begin{subfigure}[t]{0.49\linewidth}
		\centering
		\includegraphics[width=\linewidth]{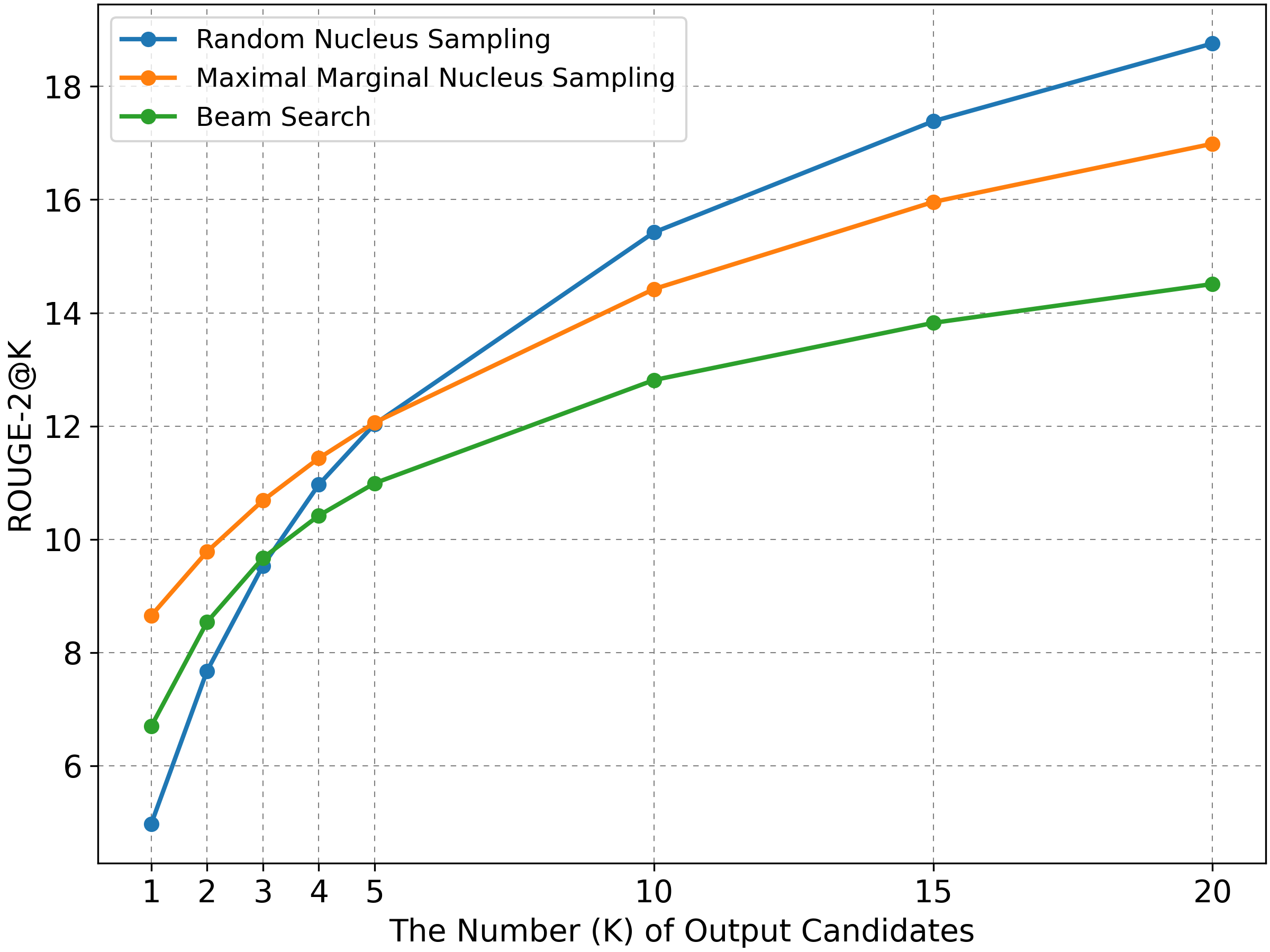}
		\caption{ROUGE-2@$K$}
		\label{rouge2}
	\end{subfigure}
	\hfill
	\begin{subfigure}[t]{0.49\linewidth}
		\centering
		\includegraphics[width=\linewidth]{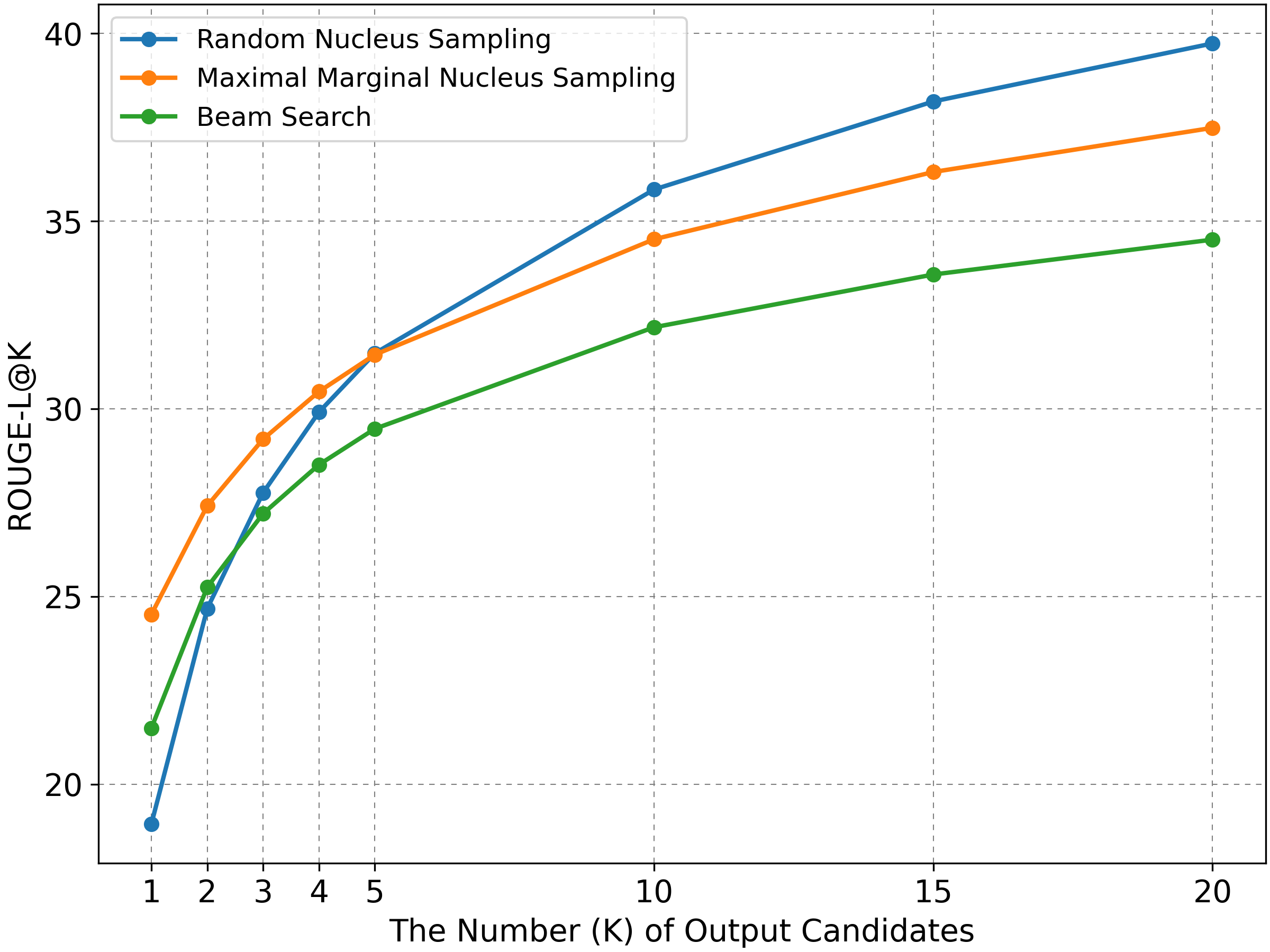}
		\caption{ROUGE-L@$K$}
		\label{rougel}
	\end{subfigure}
	\caption{Automatic evaluation results of equipping the fine-tuned CodeT5 model with three sampling strategies.}
	\label{sampling_result}
\end{figure*}

\subsection{RQ-1: Does our approach outperform state-of-the-art baselines under automatic evaluation?}
\label{rq1}

\noindent \textbf{Methods: }We compare M\textsubscript{3}NSCT5 with six state-of-the-art baselines on the four automatic evaluation metrics (i.e., BLEUS-4, ROUGE-1, ROUGE-2, and ROUGE-L). Since we propose to overcome the \textit{ambiguity} issue by sampling multiple times, we also experiment with the number of outputs $K$=3 and $K$=5. All the models (except for BM25) are trained on the whole training set of our $D_{so}$ dataset that covers eight PLs and tested on individual subsets of different PLs. The experimental results of RQ-1 are shown in Table \ref{baselines_k_equal_1} and Table \ref{baselines_k_equal_35}.

\noindent \textbf{Results: } Based on the results, we can conclude that M\textsubscript{3}NSCT5 achieves the best performance under automatic evaluation, outperforming all the baseline models. Specifically, we have the following findings:

\begin{enumerate}[(1)]
	\item According to Table \ref{baselines_k_equal_1}, when all the models output only one candidate (i.e., $K$=1), M\textsubscript{3}NSCT5 could achieve the best performance.
	Among all the baselines, the retrieval method BM25 has the worst performance. The LSTM-based Code2Que outperforms BM25 by a large margin but is no match for large pre-trained Transformer models, which aligns with the results of the previous study~\cite{liu2022sotitle}. We also find that BART, CCBERT, and SOTitle share similar results on different PL subsets, where all of them are worse than PLBART. 
    We attribute the good performance of PLBART to its generation-oriented denoising objectives and code-related corpus for pre-training. Furthermore, M\textsubscript{3}NSCT5 outperforms PLBART by 3.3\%, 8.8\%, 16.5\%, and 7.9\% in terms of BLEUS-4, ROUGE-1, ROUGE-2, and ROUGE-L on average of different PL subsets.

	\item According to Table \ref{baselines_k_equal_35}, when all the models output multiple candidates ($K$=3 and $K$=5), M\textsubscript{3}NSCT5 could also significantly improve the performance as well as outperform other baselines.
	We choose BM25 for comparison because returning multiple candidates for a query is common in the field of information retrieval. We also compare with PLBART, which shares a similar model architecture with CodeT5 and has the most competitive results when $K$=1. 
	
	As shown in Table \ref{baselines_k_equal_35}, increasing the output number $K$ (i.e., offering more title candidates for the developers to choose from) boosts the performance of all models by a large margin. 
    In particular, when $K$ changes from 1 to 3, M\textsubscript{3}NSCT5 performs 21.5\%, 18.9\%, 23.7\%, and 19.1\% better in terms of BLEUS-4, ROUGE-1, ROUGE-2, and ROUGE-L on average of different PL subsets. As for the baselines, BM25 remains the worst performance. Though PLBART achieves acceptable results, M\textsubscript{3}NSCT5 still outperforms it by 4.7\%, 8.6\%, 12\%, and 8.1\% in terms of BLEUS-4, ROUGE-1, ROUGE-2, and ROUGE-L on average when $K$=3, and by 4.9\%, 8.6\%, 11.7\%, and 7.9\% in terms of BLEUS-4, ROUGE-1, ROUGE-2, and ROUGE-L on average when $K$=5.

	\item According to the sub-tables of different PLs in Table \ref{baselines_k_equal_1} and Table \ref{baselines_k_equal_35},  M\textsubscript{3}NSCT5 achieves excellent performance on every PL subset. Surprisingly, the results on less popular PLs like \textit{C}, \textit{Ruby}, and \textit{Go} are totally comparable to the dominant ones.
	This finding also applies to other baselines, where models never pre-trained on code-related corpus like Code2Que, BART and SOTitle can perform well on all subsets. Though PLBART has only been pre-trained on \textit{Python} and \textit{Java} corpus, it can achieve quite competitive results to our model on other PLs. All the evidence indicates that models can benefit from fine-tuning on the joint dataset of different PLs, and it is applicable to introduce new PLs with less sufficient data to the SO title generation task.
\end{enumerate}

\begin{tcolorbox}
	Answer to RQ-1: Our proposed M\textsubscript{3}NSCT5 can outperform state-of-the-art baselines on automatic evaluation by generating titles of higher quality under different experimental settings.
\end{tcolorbox}

\begin{table*}
	\centering
	\caption{Human evaluation results of M\textsubscript{3}NSCT5, PLBART, and BM25 on three programming languages when $K=3$. 
	S-1, S-2, and S-3 represent the percentage of samples rated to score 1, 2, and 3. S-Avg represents the average score of title candidates. The numbers in the table are the mean values of two participants in the same group.
	} \label{human_eval}
	\begin{tabular}{@{\ \ }cccccccccccccc@{\ \ }}
		\toprule
        \multirow{3}{*}{Language} & \multirow{3}{*}{Criteria(@3)} & \multicolumn{4}{c}{M\textsubscript{3}NSCT5} &
        \multicolumn{4}{c}{PLBART} &
        \multicolumn{4}{c}{BM25} \\
        \cmidrule(lr){3-6}
        \cmidrule(lr){7-10}
        \cmidrule(lr){11-14}
        & & S-1 & S-2 & S-3 & S-Avg & S-1 & S-2 & S-3 & S-Avg & S-1 & S-2 & S-3 & S-Avg \\
        \midrule[0.6pt]
		\multirow{3}{*}{Python}
		& Readability 
		& - & 23\% & 77\%  & 2.77
		& - & 24\% & 76\%  & 2.76
		& - & 13\% & 87\%  & \textbf{2.87} \\
		& Diversity 
		& 3\% & 33\% & 64\% & \textbf{2.61}
		& 17\% & 49\% & 34\% & 2.17
		& 5\% & 35\% & 60\% & 2.55 \\ 
		& Correlation 
		& 11\% & 52\% & 37\% & \textbf{2.26}
		& 19\% & 49\% & 32\% & 2.13
		& 68\% & 32\% & - & 1.32 \\ \midrule[0.2pt]
		\multirow{3}{*}{Go}
		& Readability 
		& - & 39\% & 61\% & 2.61
		& - & 38\% & 62\% & 2.62
		& - & 16\% & 84\% & \textbf{2.84} \\
		& Diversity 
		& 4\% & 39\% & 57\% & 2.53
		& 21\% & 51\% & 28\% & 2.07
		& 3\% & 37\% & 60\% & \textbf{2.57}\\ 
		& Correlation 
		& 12\% & 55\% & 33\% & \textbf{2.21}
		& 19\% & 54\% & 27\% & 2.08
		& 71\% & 29\% & - & 1.29 \\ \midrule[0.2pt]
		\multirow{3}{*}{Java}
		& Readability 
		& - & 31\% & 69\% & 2.69
		& - & 32\% & 68\% & 2.68
		& - & 14\% & 86\% & \textbf{2.86} \\
		& Diversity 
		& 3\% & 36\% & 61\% & \textbf{2.58}
		& 18\% & 49\% & 33\% & 2.15
		& 4\% & 36\% & 60\% & 2.56 \\ 
		& Correlation 
		& 10\% & 56\% & 34\% & \textbf{2.24}
		& 19\% & 51\% & 30\% & 2.11
		& 69\% & 31\% & - & 1.31 \\\bottomrule
	\end{tabular}
\end{table*}

\subsection{RQ-2: How effective is our maximal marginal multiple nucleus sampling?}
\label{rq3}

\noindent \textbf{Methods: }Remaining the already fine-tuned CodeT5 unchanged, we compare the performance of the three sampling strategies when $K$ varies from 1 to 20. First, we use the vanilla Beam Search (BS) method, which ranks the generated candidates by the combinatorial probability of their tokens. We set the beam size to 20 and select the top $K$ candidates as output. Second, we repeat Random Nucleus Sampling (RNS) $K$ times, then take the sampled candidates as output. The third is our proposed strategy, which mainly performs Maximal Marginal Nucleus Sampling (MMNS). The experimental results of RQ-2 are shown in Figure \ref{sampling_result}, where the sub-figures individually demonstrate the performance of the three sampling strategies on each automatic evaluation metric, averaged on eight PL subsets.

\noindent \textbf{Results: } From the results, we can easily find the superiority of our sampling strategy, especially when $K\leq 5$. Specifically, we have the following findings:
\begin{enumerate}[(1)]
    \item The performance of RNS has a large fluctuation range and is highly sensitive to the value of $K$. While BS has a more moderate performance improvement than RNS when $K$ increases. Considering the user scenario of the SO title generation task, it is applicable when $K$ takes small values, e.g., recommending at most 5 title candidates for the developer. When $K\leq 2$, BS outperforms RNS, indicating the titles generated with higher probability are of better quality. When $3\leq K\leq 5$, BS performs worse than RNS, showing that ranking titles purely on probability could damage the trait of diversity.

    \item Our MMNS strategy takes advantage of nucleus sampling and maximal marginal ranking to increase the diversity of output titles. It also incorporates self-consistency voting to obtain the most promising title with higher quality. As a result, when $K$=1, MMNS outperforms BS (and RNS) by 9\%, 14.6\%, 29.2\%, and 14.1\% (by 18.2\%, 27.4\%, 74\%, and 29.5\%) in terms of BLEUS-4, ROUGE-1, ROUGE-2, and ROUGE-L. When $K$=3, MMNS outperforms RNS (and BS) by 3.5\%, 4.5\%, 12.1\%, and 5.1\% (by 2.3\%, 7.9\%, 10.5\%, and 7.3\%) in terms of BLEUS-4, ROUGE-1, ROUGE-2, and ROUGE-L. When $K$=5, MMNS is still comparable to RNS.
\end{enumerate}

\begin{tcolorbox}
	Answer to RQ-2: Our sampling strategy can improve the quality and diversity of generated titles, especially when $K\leq 5$, making it more suitable and effective for the SO title generation task.
\end{tcolorbox}

\subsection{RQ-3: What is the performance of our approach under human evaluation?}
\label{rq2}
\noindent \textbf{Methods: }As introduced in Section \ref{human_evaluation}, we recruit six students for human evaluation. Participants are divided into three groups and are required to score the titles generated by M\textsubscript{3}NSCT5, PLBART, and BM25 on three PLs. Finally, we take the average score of each two participants in the same group and report the results by different PLs. We also employ Cohen's Kappa~\cite{cohen1960coefficient} to measure the agreement between the two participants in each group with respect to the languages, criteria, and models.
The main evaluation results of RQ-3 are shown in Table \ref{human_eval} and the Cohen's Kappa statistics are summarized in Table \ref{tab:agreement_explain} and Table \ref{tab:agreement_value}. In addition, we put some examples in Table \ref{tab:examples} to demonstrate the quality and diversity of the titles generated by the three approaches. \\

\noindent \textbf{Results: } From the results, we can find that M\textsubscript{3}NSCT5 has the best performance in terms of \textit{Diversity} and \textit{Correlation} compared with the other two approaches. And the performance of M\textsubscript{3}NSCT5 is stable among different programming languages, even for the low-resource $Go$ subset. Moreover, the participants have a substantial or nearly perfect agreement according to Cohen's Kappa statistics, which validates the trustworthiness of our human evaluation. Specifically, we have the following findings:
\begin{enumerate}[(1)]
    \item In terms of the \textit{Readability} criterion, BM25 achieves the best performance because it just returns the human-written titles retrieved from the training set. Besides, both M\textsubscript{3}NSCT5 and PLBART can achieve a competitive score $\geq 2.6$ on three PLs, indicating the capability of pre-trained models on natural language generation. The examples in Table \ref{tab:examples} also demonstrate the good readability of generated titles.
    
    \item In terms of the \textit{Diversity} criterion, both M\textsubscript{3}NSCT5 and BM25 have good results, having an average number of distinct titles $\geq 2.5$ when $K=3$. BM25 performs well because there are almost no duplicate posts in the training set. The excellent performance of M\textsubscript{3}NSCT5 should attribute to our elaborate sampling strategy that maximizes the difference between the output titles. In contrast, the poor performance of PLBART on \textit{Diversity} should blame on the beam search sampling method. From the examples in Table \ref{tab:examples}, we can find that the titles generated by PLBART have higher lexical and semantic overlap than our approach.
    
    \item As for the \textit{Correlation} criterion, M\textsubscript{3}NSCT5 outperforms PLBART and BM25, having around 90\% samples relevant to or exactly matching the key points of original posts. It shows the feasibility of inferring user intents from code snippets. We may attribute the excellent performance of M\textsubscript{3}NSCT5 to the initial choice of high-quality titles based on self-consistency when performing the maximal marginal ranking. We can see from Table \ref{tab:examples} that BM25 is totally off the point and even recommends \textit{PHP} posts for the \textit{Go} code snippets because of their high lexical overlap, indicating the limitations of the retrieval method.
    The titles generated by PLBART are relevant to the posts but still missing the points. At the same time, M\textsubscript{3}NSCT5 shows a good ability to understand the code snippets and generate diverse title candidates, with the best candidate closely related to the post.

\end{enumerate}

\begin{tcolorbox}
	Answer to RQ-3: Our approach shows a strong ability to generate post titles with high quality and diversity on different programming languages under human evaluation.
\end{tcolorbox}

\begin{table}[t]
    \centering
    \caption{The interpretation of Cohen's Kappa agreement.}
    \label{tab:agreement_explain}
    \scalebox{1.1}{
    	\begin{tabular}{@{\ \ }cc@{\ \ }}
    		\toprule
            Cohen's Kappa & Interpretation \\
            \midrule[0.6pt]
            0\% & No agreement \\
    		1\% $\sim$ 20\% & Slight agreement \\
    		21\% $\sim$ 40\% & Fair agreement \\
    		41\% $\sim$ 60\% & Moderate agreement \\
    		61\% $\sim$ 80\% & Substantial agreement \\
    		81\% $\sim$ 99\% & Near perfect agreement \\
    		100\% & Perfect agreement \\ \bottomrule
    	\end{tabular}
	}
\end{table}

\begin{table}[t]
    \centering
    \caption{The agreement values of human evaluation results.}
    \label{tab:agreement_value}
    \scalebox{0.95}{
    	\begin{tabular}{@{\ \ }ccccc@{\ \ }}
    		\toprule
            Language & Criteria(@3) & M\textsubscript{3}NSCT5 & PLBART & BM25 \\
            \midrule[0.6pt]
    		\multirow{3}{*}{Python}
    		& Readability 
    		& 83.1\% & 78.3\% & 73.7\% \\
    		& Diversity 
    		& 79.4\% & 83.9\% & 84.5\% \\ 
    		& Correlation 
    		& 79.6\% & 84.0\% & 81.8\% \\ \midrule[0.2pt]
    		\multirow{3}{*}{Go}
    		& Readability 
    		& 79.2\% & 74.9\% & 85.2\% \\
    		& Diversity 
    		& 81.0\% & 77.6\% & 80.3\% \\ 
    		& Correlation 
    		& 82.7\% & 76.9\% & 76.0\% \\ \midrule[0.2pt]
    		\multirow{3}{*}{Java}
    		& Readability 
    		& 76.9\% & 81.8\% & 83.4\% \\
    		& Diversity 
    		& 80.1\% & 77.7\% & 84.4\% \\ 
    		& Correlation 
    		& 85.8\% & 80.6\% & 86.0\% \\\bottomrule
    	\end{tabular}
	}
\end{table}

\begin{table*}[p]
	\centering
	\caption{Example input code snippets and the titles generated by M\textsubscript{3}NSCT5, PLBART, and BM25. URL links of the original posts and those retrieved by BM25 are listed. }\label{tab:examples}
	\resizebox{0.99\textwidth}{!}{
	\begin{tabular}{@{\ \ }l|l@{\ }}
		\toprule
		Code Snippets  & Titles \\ \midrule[0.6pt]
		\multirow{14}{7.3cm}{
			\textbf{Example --- Python Language:} \\
			letter\_list = [\textcolor{OliveGreen}{'a'},\textcolor{OliveGreen}{'d'},\textcolor{OliveGreen}{'o'},\textcolor{OliveGreen}{'m'},\textcolor{OliveGreen}{'s'}] \\
			$>>>$~df \\
			ID~~~~WORD \\
			\textcolor{RawSienna}{1}~~~~~\textcolor{OliveGreen}{'yellow'} \\
			\textcolor{RawSienna}{2}~~~~~\textcolor{OliveGreen}{'orange'} \\
			\textcolor{RawSienna}{3}~~~~~\textcolor{OliveGreen}{'green'} \\
			\textcolor{RawSienna}{4}~~~~~\textcolor{OliveGreen}{'blue'} \\
			$>>>$~expected output \\
			ID~~~~WORD \\
			\textcolor{RawSienna}{3}~~~~~\textcolor{OliveGreen}{'green'} \\
			\textcolor{RawSienna}{4}~~~~~\textcolor{OliveGreen}{'blue'} \\
		}
		& \textbf{\scriptsize{Origin}}: 
		 \\
		& \underline{filter dataframe for words which do not contain any of the letters in a list}  \\
		& \textbf{\scriptsize{M\textsubscript{3}NSCT5}}: \\
		& - how to filter rows in pandas dataframe based on a list of letters \\
		& - pandas - replace rows with values of list from another list \\
		& - pandas : groupby by letters of string \\
		& \textbf{\scriptsize{PLBART}}: \\
		& - use pandas to group words from a list \\
		& - how to group elements of a list in pandas \\
		& - how to group by a second list in pandas \\
		& \textbf{\scriptsize{BM25}}: \\
		& - pandas merge creates unwanted duplicate entries 
		\\
		& - using nested loops in order to group data in a csv 
		\\
		& - sort a list with known values before unknown values 
		\\
		\midrule[0.4pt]

		\multirow{14}{7.3cm}{
			\textbf{Example --- Java Language:} \\
			\textcolor{NavyBlue}{try}(\textcolor{RawSienna}{Scanner} \textcolor{OliveGreen}{scan} = \textcolor{NavyBlue}{new} \textcolor{RawSienna}{Scanner}(System.in)) \{ \\
			\hspace{4mm}\textcolor{RawSienna}{String} \textcolor{OliveGreen}{s} = scan.nextLine();\\
			\hspace{4mm}System.out.println(s);\\
			\hspace{4mm}scan = \textcolor{RawSienna}{null};\\
			\}
		}
		& \textbf{\scriptsize{Origin}}:
		\\
		& \underline{try with resource why cannot modify resource} \\
		& \textbf{\scriptsize{M\textsubscript{3}NSCT5}}: \\
		& - how to close a scanner in a try-with-resources block \\
		& - why is the scanner null after the first line \\
		& - java scanner.nextline() not working as expected \\
		& \textbf{\scriptsize{PLBART}}: \\
		& - will it be a nullpointerexception in try-with-resource \\
		& - java scanner() throws nullpointerexception \\
		& - how do I prevent a nullpointerexception from being thrown \\
		& \textbf{\scriptsize{BM25}}: \\
		& - copying characters in a string
		\\
		& - java how to parse for words in a string for a specific word
		\\
		& - how to terminate scanner when input is complete
		\\
		\midrule[0.4pt]
		
        \multirow{14}{7.3cm}{
			\textbf{Example --- Go Language:} \\
   			\{\{\textcolor{NavyBlue}{range} \textcolor{OliveGreen}{\$m} := .messages\}\}  \\    
			\hspace{4mm}<\textcolor{RawSienna}{div}>Message subject: \{\{\$m.Subject\}\}<\textcolor{RawSienna}{/div}> \\
			\hspace{4mm}\{\{\textcolor{OliveGreen}{\$lastMsg} := \$m\}\}\\ \{\{\textcolor{NavyBlue}{end}\}\} \\
			<\textcolor{RawSienna}{div}>\\
			\hspace{4mm}The last message's subject: \{\{\$lasMsg.Subject\}\}\\
			<\textcolor{RawSienna}{/div}> \\
			~\\
			undefined variable \textcolor{OliveGreen}{``\$lastMsg''} \\
			~\\
			unexpected \textcolor{OliveGreen}{``:=''} in operand \\
		}
		& \textbf{\scriptsize{Origin}}:
		\\
		& \underline{how to use template variable outside a loop} \\
		& \textbf{\scriptsize{M\textsubscript{3}NSCT5}}: \\
		& - how to access a variable in a range in go template \\
		& - golang nested loop variable not found \\
		& - why does \{\{\$variable\}\} not work in this case \\
		& \textbf{\scriptsize{PLBART}}: \\
		& - go range variables in html template \\
		& - go templates, use variable in range \\
		& - accessing a slice value inside a range in go \\
		& \textbf{\scriptsize{BM25}}: \\
		& - rails: form in partial for new nested resource
		\\
		& - how to use visual studio - generated async wcf calls
		\\
		& - how do i secure this php script 
		\\
		\bottomrule
	\end{tabular}
}
\end{table*}

\section{Related Work}
\label{related work}
A Stack Overflow post usually consists of three parts: body, title, and tags. Previous studies on the tag recommendation task demonstrated that one could utilize the recommended tags for post retrieval, similar to our motivation. But the discrete tags are more suitable for post classification than serving as search queries. In comparison, a coherent and informative post title can better help developers understand the problem and search for related posts. Generating post titles from code snippets can be seen as a specialized PL-to-NL translation task. Another closely related task is code summarization, an emerging research direction in software engineering and natural language processing. This section introduces the previous studies of post title generation, tag recommendation, and the recent advances in code summarization.

\paragraph{Title Generation:}
Gao et al.~\cite{gao2020generating} first proposed the SO title generation task to help improve the quality of poorly written question posts. They trained an LSTM network that equips with the copy~\cite{See2017GetTT} and coverage~\cite{Tu2016ModelingCF} mechanisms to generate titles from mined code snippets. Later, Zhang et al.~\cite{zhang2022improving} and Liu et al.~\cite{liu2022sotitle} found that taking advantage of both code snippets and text descriptions in the post body 
could significantly improve the quality of generated titles. Though utilizing the natural language descriptions could reduce the ambiguity of the context, it is less helpful when developers cannot provide good question descriptions. Therefore, in this study, we focus on the application scenario in which only code snippets are available. We propose M\textsubscript{3}NSCT5 to further improve the quality and diversity of generated titles compared with previous approaches.

\paragraph{Tag Recommendation:}
Post tags are vital for Stack Overflow, which are helpful in organizing relevant posts. Nevertheless, poorly chosen tags may cause severe redundancy over time. To tackle this challenge, early studies~\cite{Xia2013TagRI, Wang2014EnTagRecAE, Wang2015TagCombineRT, Zhou2017ScalableTR, Liu2018FastTagRecFT} proposed to automatically recommend tags with the given post body, title, and user profile through feature extraction and similarity-based methods. Recently, Zhou et al.~\cite{zhou2019deep} and Xu et al.~\cite{Xu2021Post2VecLD} introduced end-to-end deep learning models for this task, which could achieve better performance. Moreover, Devine et al.~\cite{Devine2022UnsupervisedEM} and He et al.~\cite{He2022PTM4TagST} proposed to take advantage of pre-trained models for further improvement. However, it remains unexplored to recommend post tags with only code snippets. We believe tag recommendation and title generation models can be combined for post retrieval, which is a direction of our future work.

\paragraph{Code Summarization:}
The Code summarization task is to generate readable descriptions of the given program, aiming to save the effort of developers on program comprehension. With the emergence of large-scale NL-PL bi-modal datasets, it becomes feasible to train deep Transformer models to generate high-quality code summaries. Ahmad et al.~\cite{ahmad2020transformer} first employed the Transformer encoder-decoder model to handle the long-range dependencies between code tokens and outperformed previous LSTM-based models by a large margin. Then, a number of follow-up studies~\cite{tang2021ast, Tang2022ASTTransCS, Shi2021CASTEC, guo2022modeling, Zhou2022AutomaticSC, Li2021SeCNNAS, Cheng2022KeywordguidedAC, Zhou2022SummarizingSC} proposed to incorporate the structural information by parsing the source code into abstract syntax trees or control flow graphs to improve the performance. Since code snippets extracted from SO posts are always problematic, we cannot apply static parsing techniques to get the syntax trees or control flow graphs. Therefore, as mentioned in Section \ref{data_preparation}, we try to reserve the structural information by keeping the white spaces and line breaks in code snippets.  Furthermore, some other studies~\cite{Lu2021CodeXGLUEAM, feng2020codebert, ahmad2021unified, Wang2021CodeT5IU, Guo2022UniXcoderUC, fried2022incoder} proposed to utilize the large-scale unlabelled data and self-supervised learning to pre-train models through self-supervised objectives and achieved state-of-the-art results on code summarization benchmarks. Motivated by the good performance of pre-training, we employ the pre-trained CodeT5 model as our backbone. 

\section{Threats to Validity}
\label{threats to validity}
This section reveals the potential threats that may affect the reproduction of our experiments and the validation of our results.

\textbf{The threats to internal validity} mainly relate to the implementation of baseline models. For CCBERT and SOTitle that already have released source code and model checkpoints, we keep their default hyper-parameters unchanged in our experiments. For BM25, Code2Que, BART, and PLBART, which have no off-the-shelf implementations, we take advantage of the widely used libraries (i.e., Elasticsearch, OpenNMT, and transformers) for reproduction and tune their hyper-parameters to the best on our dataset. In this way, we make sure the comparison between our model and the baselines is fair. And we release our implementations of the baselines to facilitate future studies.

\textbf{The threats to external validity} mainly relate to the construction of our dataset. We have tried our best to ensure the quality of our dataset. First, we utilize the already processed dataset SOTorrent to avoid potential bugs when extracting code snippets from the post body. Second, we only include the filtered high-quality posts in the dataset to reduce the noise of training and test data. Third, our dataset covers eight programming languages, including the minorities (\textit{Ruby} and \textit{GO}), which would better test the generalization ability of models. Moreover, we split the train and test sets in chronological order to fit real-world scenarios. We also release our dataset for validation and reproduction.

\textbf{The threats to construct validity} mainly relate to the evaluation methods. Though BLEU and ROUGE are the most popular evaluation metrics for generation tasks, measuring the quality of the generated content remains an open challenge. In the SO title generation task, there is no golden title for a given post, which makes it unfair to judge the quality of generated titles by comparing them with the only reference title. Therefore, we perform an additional human evaluation to show the intuitive quality of generated titles. To perform a comprehensive study, we invite six participants to evaluate the posts covering three programming languages.

\section{Conclusion and Future Work}
\label{conclusion and future work}
In this paper, we proposed M\textsubscript{3}NSCT5, a novel approach to automatically generate Stack Overflow post titles from the given code snippets, which can help non-native English speakers or inexperienced developers improve their poorly written question posts. Combining the pre-trained CodeT5 model and the maximal marginal multiple nucleus sampling strategy, M\textsubscript{3}NSCT5 can generate high-quality and diverse title candidates for the developers to choose from.
To validate the effectiveness of our approach, we have built a large-scale dataset with 890,000 posts covering eight programming languages and choose six state-of-the-art baselines for comparison. 
We performed extensive experiments to demonstrate the superiority of our approach, including an automatic evaluation on the BLEU and ROUGE metrics and a human evaluation using the \textit{Readability}, \textit{Correlation}, and \textit{Diversity} criteria. Results showed that M\textsubscript{3}NSCT5 outperforms all the baseline methods by a significant margin and has great potential for real-world application.

For future work, we plan to further incorporate more powerful pre-trained language models and tag recommendation methods to improve the title generation task performance. Moreover, we plan to deploy our model as web services so that real-world developers from Stack Overflow could benefit from our work and produce valuable user feedback for future improvement.

\bibliographystyle{cas-model2-names}

\bibliography{main}

\begin{thebibliography}{56}
\expandafter\ifx\csname natexlab\endcsname\relax\def\natexlab#1{#1}\fi
\providecommand{\url}[1]{\texttt{#1}}
\providecommand{\href}[2]{#2}
\providecommand{\path}[1]{#1}
\providecommand{\DOIprefix}{doi:}
\providecommand{\ArXivprefix}{arXiv:}
\providecommand{\URLprefix}{URL: }
\providecommand{\Pubmedprefix}{pmid:}
\providecommand{\doi}[1]{\href{http://dx.doi.org/#1}{\path{#1}}}
\providecommand{\Pubmed}[1]{\href{pmid:#1}{\path{#1}}}
\providecommand{\bibinfo}[2]{#2}
\ifx\xfnm\relax \def\xfnm[#1]{\unskip,\space#1}\fi
\bibitem[{Ahmad et~al.(2020)Ahmad, Chakraborty, Ray and
  Chang}]{ahmad2020transformer}
\bibinfo{author}{Ahmad, W.}, \bibinfo{author}{Chakraborty, S.},
  \bibinfo{author}{Ray, B.}, \bibinfo{author}{Chang, K.W.},
  \bibinfo{year}{2020}.
\newblock \bibinfo{title}{A transformer-based approach for source code
  summarization}, in: \bibinfo{booktitle}{Proceedings of the 58th Annual
  Meeting of the Association for Computational Linguistics}, pp.
  \bibinfo{pages}{4998--5007}.
\bibitem[{Ahmad et~al.(2021)Ahmad, Chakraborty, Ray and
  Chang}]{ahmad2021unified}
\bibinfo{author}{Ahmad, W.}, \bibinfo{author}{Chakraborty, S.},
  \bibinfo{author}{Ray, B.}, \bibinfo{author}{Chang, K.W.},
  \bibinfo{year}{2021}.
\newblock \bibinfo{title}{Unified pre-training for program understanding and
  generation}, in: \bibinfo{booktitle}{Proceedings of the 2021 Conference of
  the North American Chapter of the Association for Computational Linguistics:
  Human Language Technologies}, pp. \bibinfo{pages}{2655--2668}.
\bibitem[{Baltes et~al.(2018)Baltes, Dumani, Treude and
  Diehl}]{baltes2018sotorrent}
\bibinfo{author}{Baltes, S.}, \bibinfo{author}{Dumani, L.},
  \bibinfo{author}{Treude, C.}, \bibinfo{author}{Diehl, S.},
  \bibinfo{year}{2018}.
\newblock \bibinfo{title}{Sotorrent: Reconstructing and analyzing the evolution
  of stack overflow posts}, in: \bibinfo{booktitle}{Proceedings of the 15th
  international conference on mining software repositories}, pp.
  \bibinfo{pages}{319--330}.
\bibitem[{Chatterjee et~al.(2020)Chatterjee, Kong and
  Pollock}]{Chatterjee2020FindingHW}
\bibinfo{author}{Chatterjee, P.}, \bibinfo{author}{Kong, M.},
  \bibinfo{author}{Pollock, L.}, \bibinfo{year}{2020}.
\newblock \bibinfo{title}{Finding help with programming errors: An exploratory
  study of novice software engineers’ focus in stack overflow posts}.
\newblock \bibinfo{journal}{Journal of Systems and Software}
  \bibinfo{volume}{159}, \bibinfo{pages}{110454}.
\bibitem[{Chen et~al.(2021)Chen, Tworek, Jun, Yuan, Pinto, Kaplan, Edwards,
  Burda, Joseph, Brockman et~al.}]{chen2021evaluating}
\bibinfo{author}{Chen, M.}, \bibinfo{author}{Tworek, J.}, \bibinfo{author}{Jun,
  H.}, \bibinfo{author}{Yuan, Q.}, \bibinfo{author}{Pinto, H.P.d.O.},
  \bibinfo{author}{Kaplan, J.}, \bibinfo{author}{Edwards, H.},
  \bibinfo{author}{Burda, Y.}, \bibinfo{author}{Joseph, N.},
  \bibinfo{author}{Brockman, G.}, et~al., \bibinfo{year}{2021}.
\newblock \bibinfo{title}{Evaluating large language models trained on code}.
\newblock \bibinfo{journal}{arXiv preprint arXiv:2107.03374} .
\bibitem[{Cheng et~al.(2022)Cheng, Hu, Wei and Mo}]{Cheng2022KeywordguidedAC}
\bibinfo{author}{Cheng, W.}, \bibinfo{author}{Hu, P.}, \bibinfo{author}{Wei,
  S.}, \bibinfo{author}{Mo, R.}, \bibinfo{year}{2022}.
\newblock \bibinfo{title}{Keyword-guided abstractive code summarization via
  incorporating structural and contextual information}.
\newblock \bibinfo{journal}{Information and Software Technology}
  \bibinfo{volume}{150}, \bibinfo{pages}{106987}.
\bibitem[{Cobbe et~al.(2021)Cobbe, Kosaraju, Bavarian, Hilton, Nakano, Hesse
  and Schulman}]{cobbe2021training}
\bibinfo{author}{Cobbe, K.}, \bibinfo{author}{Kosaraju, V.},
  \bibinfo{author}{Bavarian, M.}, \bibinfo{author}{Hilton, J.},
  \bibinfo{author}{Nakano, R.}, \bibinfo{author}{Hesse, C.},
  \bibinfo{author}{Schulman, J.}, \bibinfo{year}{2021}.
\newblock \bibinfo{title}{Training verifiers to solve math word problems}.
\newblock \bibinfo{journal}{arXiv preprint arXiv:2110.14168} .
\bibitem[{Cohen(1960)}]{cohen1960coefficient}
\bibinfo{author}{Cohen, J.}, \bibinfo{year}{1960}.
\newblock \bibinfo{title}{A coefficient of agreement for nominal scales}.
\newblock \bibinfo{journal}{Educational and psychological measurement}
  \bibinfo{volume}{20}, \bibinfo{pages}{37--46}.
\bibitem[{Devine and Blincoe(2022)}]{Devine2022UnsupervisedEM}
\bibinfo{author}{Devine, P.}, \bibinfo{author}{Blincoe, K.},
  \bibinfo{year}{2022}.
\newblock \bibinfo{title}{Unsupervised extreme multi label classification of
  stack overflow posts}.
\newblock \bibinfo{journal}{2022 IEEE/ACM 1st International Workshop on Natural
  Language-Based Software Engineering (NLBSE)} , \bibinfo{pages}{1--8}.
\bibitem[{Feng et~al.(2020)Feng, Guo, Tang, Duan, Feng, Gong, Shou, Qin, Liu,
  Jiang et~al.}]{feng2020codebert}
\bibinfo{author}{Feng, Z.}, \bibinfo{author}{Guo, D.}, \bibinfo{author}{Tang,
  D.}, \bibinfo{author}{Duan, N.}, \bibinfo{author}{Feng, X.},
  \bibinfo{author}{Gong, M.}, \bibinfo{author}{Shou, L.}, \bibinfo{author}{Qin,
  B.}, \bibinfo{author}{Liu, T.}, \bibinfo{author}{Jiang, D.}, et~al.,
  \bibinfo{year}{2020}.
\newblock \bibinfo{title}{Codebert: A pre-trained model for programming and
  natural languages}, in: \bibinfo{booktitle}{Findings of the Association for
  Computational Linguistics: EMNLP 2020}, pp. \bibinfo{pages}{1536--1547}.
\bibitem[{Fried et~al.(2022)Fried, Aghajanyan, Lin, Wang, Wallace, Shi, Zhong,
  Yih, Zettlemoyer and Lewis}]{fried2022incoder}
\bibinfo{author}{Fried, D.}, \bibinfo{author}{Aghajanyan, A.},
  \bibinfo{author}{Lin, J.}, \bibinfo{author}{Wang, S.},
  \bibinfo{author}{Wallace, E.}, \bibinfo{author}{Shi, F.},
  \bibinfo{author}{Zhong, R.}, \bibinfo{author}{Yih, W.t.},
  \bibinfo{author}{Zettlemoyer, L.}, \bibinfo{author}{Lewis, M.},
  \bibinfo{year}{2022}.
\newblock \bibinfo{title}{Incoder: A generative model for code infilling and
  synthesis}.
\newblock \bibinfo{journal}{arXiv preprint arXiv:2204.05999} .
\bibitem[{Gao et~al.(2020)Gao, Xia, Grundy, Lo and Li}]{gao2020generating}
\bibinfo{author}{Gao, Z.}, \bibinfo{author}{Xia, X.}, \bibinfo{author}{Grundy,
  J.}, \bibinfo{author}{Lo, D.}, \bibinfo{author}{Li, Y.F.},
  \bibinfo{year}{2020}.
\newblock \bibinfo{title}{Generating question titles for stack overflow from
  mined code snippets}.
\newblock \bibinfo{journal}{ACM Transactions on Software Engineering and
  Methodology (TOSEM)} \bibinfo{volume}{29}, \bibinfo{pages}{1--37}.
\bibitem[{Gao et~al.(2022)Gao, Xia, Lo, Grundy, Zhang and Xing}]{Gao2022IKW}
\bibinfo{author}{Gao, Z.}, \bibinfo{author}{Xia, X.}, \bibinfo{author}{Lo, D.},
  \bibinfo{author}{Grundy, J.C.}, \bibinfo{author}{Zhang, X.},
  \bibinfo{author}{Xing, Z.}, \bibinfo{year}{2022}.
\newblock \bibinfo{title}{I know what you are searching for: Code snippet
  recommendation from stack overflow posts}.
\newblock \bibinfo{journal}{ACM Transactions on Software Engineering and
  Methodology} \DOIprefix\doi{10.1145/3550150}.
\bibitem[{Guo et~al.(2022a)Guo, Lu, Duan, Wang, Zhou and
  Yin}]{Guo2022UniXcoderUC}
\bibinfo{author}{Guo, D.}, \bibinfo{author}{Lu, S.}, \bibinfo{author}{Duan,
  N.}, \bibinfo{author}{Wang, Y.}, \bibinfo{author}{Zhou, M.},
  \bibinfo{author}{Yin, J.}, \bibinfo{year}{2022}a.
\newblock \bibinfo{title}{Unixcoder: Unified cross-modal pre-training for code
  representation}.
\newblock \bibinfo{journal}{arXiv preprint arXiv:2203.03850} .
\bibitem[{Guo et~al.(2022b)Guo, Liu, Wan, Li and Zhou}]{guo2022modeling}
\bibinfo{author}{Guo, J.}, \bibinfo{author}{Liu, J.}, \bibinfo{author}{Wan,
  Y.}, \bibinfo{author}{Li, L.}, \bibinfo{author}{Zhou, P.},
  \bibinfo{year}{2022}b.
\newblock \bibinfo{title}{Modeling hierarchical syntax structure with triplet
  position for source code summarization}, in: \bibinfo{booktitle}{Proceedings
  of the 60th Annual Meeting of the Association for Computational Linguistics
  (Volume 1: Long Papers)}, pp. \bibinfo{pages}{486--500}.
\bibitem[{He et~al.(2022)He, Xu, Yang, Han, Yang and Lo}]{He2022PTM4TagST}
\bibinfo{author}{He, J.}, \bibinfo{author}{Xu, B.}, \bibinfo{author}{Yang, Z.},
  \bibinfo{author}{Han, D.}, \bibinfo{author}{Yang, C.}, \bibinfo{author}{Lo,
  D.}, \bibinfo{year}{2022}.
\newblock \bibinfo{title}{Ptm4tag: Sharpening tag recommendation of stack
  overflow posts with pre-trained models}.
\newblock \bibinfo{journal}{2022 IEEE/ACM 30th International Conference on
  Program Comprehension (ICPC)} , \bibinfo{pages}{1--11}.
\bibitem[{Hendrycks et~al.(2021)Hendrycks, Basart, Kadavath, Mazeika, Arora,
  Guo, Burns, Puranik, He, Song et~al.}]{Hendrycks2021MeasuringCC}
\bibinfo{author}{Hendrycks, D.}, \bibinfo{author}{Basart, S.},
  \bibinfo{author}{Kadavath, S.}, \bibinfo{author}{Mazeika, M.},
  \bibinfo{author}{Arora, A.}, \bibinfo{author}{Guo, E.},
  \bibinfo{author}{Burns, C.}, \bibinfo{author}{Puranik, S.},
  \bibinfo{author}{He, H.}, \bibinfo{author}{Song, D.}, et~al.,
  \bibinfo{year}{2021}.
\newblock \bibinfo{title}{Measuring coding challenge competence with apps}.
\newblock \bibinfo{journal}{arXiv preprint arXiv:2105.09938} .
\bibitem[{Hochreiter and Schmidhuber(1997)}]{hochreiter1997long}
\bibinfo{author}{Hochreiter, S.}, \bibinfo{author}{Schmidhuber, J.},
  \bibinfo{year}{1997}.
\newblock \bibinfo{title}{Long short-term memory}.
\newblock \bibinfo{journal}{Neural computation} \bibinfo{volume}{9},
  \bibinfo{pages}{1735--1780}.
\bibitem[{Holtzman et~al.(2019)Holtzman, Buys, Du, Forbes and
  Choi}]{holtzman2019curious}
\bibinfo{author}{Holtzman, A.}, \bibinfo{author}{Buys, J.},
  \bibinfo{author}{Du, L.}, \bibinfo{author}{Forbes, M.},
  \bibinfo{author}{Choi, Y.}, \bibinfo{year}{2019}.
\newblock \bibinfo{title}{The curious case of neural text degeneration}.
\newblock \bibinfo{journal}{arXiv preprint arXiv:1904.09751} .
\bibitem[{Inala et~al.(2022)Inala, Wang, Yang, Codas, Encarnaci{\'o}n, Lahiri,
  Musuvathi and Gao}]{inala2022fault}
\bibinfo{author}{Inala, J.P.}, \bibinfo{author}{Wang, C.},
  \bibinfo{author}{Yang, M.}, \bibinfo{author}{Codas, A.},
  \bibinfo{author}{Encarnaci{\'o}n, M.}, \bibinfo{author}{Lahiri, S.K.},
  \bibinfo{author}{Musuvathi, M.}, \bibinfo{author}{Gao, J.},
  \bibinfo{year}{2022}.
\newblock \bibinfo{title}{Fault-aware neural code rankers}.
\newblock \bibinfo{journal}{arXiv preprint arXiv:2206.03865} .
\bibitem[{Kenton and Toutanova(2019)}]{kenton2019bert}
\bibinfo{author}{Kenton, J.D.M.W.C.}, \bibinfo{author}{Toutanova, L.K.},
  \bibinfo{year}{2019}.
\newblock \bibinfo{title}{Bert: Pre-training of deep bidirectional transformers
  for language understanding}, in: \bibinfo{booktitle}{Proceedings of
  NAACL-HLT}, pp. \bibinfo{pages}{4171--4186}.
\bibitem[{Khandelwal et~al.(2018)Khandelwal, He, Qi and
  Jurafsky}]{khandelwal2018sharp}
\bibinfo{author}{Khandelwal, U.}, \bibinfo{author}{He, H.},
  \bibinfo{author}{Qi, P.}, \bibinfo{author}{Jurafsky, D.},
  \bibinfo{year}{2018}.
\newblock \bibinfo{title}{Sharp nearby, fuzzy far away: How neural language
  models use context}, in: \bibinfo{booktitle}{Proceedings of the 56th Annual
  Meeting of the Association for Computational Linguistics (Volume 1: Long
  Papers)}, pp. \bibinfo{pages}{284--294}.
\bibitem[{Lewis et~al.(2020)Lewis, Liu, Goyal, Ghazvininejad, Mohamed, Levy,
  Stoyanov and Zettlemoyer}]{lewis2020bart}
\bibinfo{author}{Lewis, M.}, \bibinfo{author}{Liu, Y.}, \bibinfo{author}{Goyal,
  N.}, \bibinfo{author}{Ghazvininejad, M.}, \bibinfo{author}{Mohamed, A.},
  \bibinfo{author}{Levy, O.}, \bibinfo{author}{Stoyanov, V.},
  \bibinfo{author}{Zettlemoyer, L.}, \bibinfo{year}{2020}.
\newblock \bibinfo{title}{Bart: Denoising sequence-to-sequence pre-training for
  natural language generation, translation, and comprehension}, in:
  \bibinfo{booktitle}{Proceedings of the 58th Annual Meeting of the Association
  for Computational Linguistics}, pp. \bibinfo{pages}{7871--7880}.
\bibitem[{Li et~al.(2022)Li, Choi, Chung, Kushman, Schrittwieser, Leblond,
  Eccles, Keeling, Gimeno, Lago et~al.}]{li2022competition}
\bibinfo{author}{Li, Y.}, \bibinfo{author}{Choi, D.}, \bibinfo{author}{Chung,
  J.}, \bibinfo{author}{Kushman, N.}, \bibinfo{author}{Schrittwieser, J.},
  \bibinfo{author}{Leblond, R.}, \bibinfo{author}{Eccles, T.},
  \bibinfo{author}{Keeling, J.}, \bibinfo{author}{Gimeno, F.},
  \bibinfo{author}{Lago, A.D.}, et~al., \bibinfo{year}{2022}.
\newblock \bibinfo{title}{Competition-level code generation with alphacode}.
\newblock \bibinfo{journal}{arXiv preprint arXiv:2203.07814} .
\bibitem[{Li et~al.(2021)Li, Wu, Peng, Chen, Sun, Liu and Yu}]{Li2021SeCNNAS}
\bibinfo{author}{Li, Z.}, \bibinfo{author}{Wu, Y.}, \bibinfo{author}{Peng, B.},
  \bibinfo{author}{Chen, X.}, \bibinfo{author}{Sun, Z.}, \bibinfo{author}{Liu,
  Y.}, \bibinfo{author}{Yu, D.}, \bibinfo{year}{2021}.
\newblock \bibinfo{title}{Secnn: A semantic cnn parser for code comment
  generation}.
\newblock \bibinfo{journal}{Journal of Systems and Software}
  \bibinfo{volume}{181}, \bibinfo{pages}{111036}.
\bibitem[{Lin(2004)}]{Lin2004ROUGEAP}
\bibinfo{author}{Lin, C.Y.}, \bibinfo{year}{2004}.
\newblock \bibinfo{title}{{ROUGE}: A package for automatic evaluation of
  summaries}, in: \bibinfo{booktitle}{Text Summarization Branches Out},
  \bibinfo{publisher}{Association for Computational Linguistics}. pp.
  \bibinfo{pages}{74--81}.
\bibitem[{Lin and Och(2004)}]{Lin2004ORANGEAM}
\bibinfo{author}{Lin, C.Y.}, \bibinfo{author}{Och, F.J.}, \bibinfo{year}{2004}.
\newblock \bibinfo{title}{Orange: a method for evaluating automatic evaluation
  metrics for machine translation}, in: \bibinfo{booktitle}{Proceedings of the
  20th International Conference on Computational Linguistics}, pp.
  \bibinfo{pages}{501--507}.
\bibitem[{Liu et~al.(2018)Liu, Zhou, Yang, Liu and
  Grundy}]{Liu2018FastTagRecFT}
\bibinfo{author}{Liu, J.}, \bibinfo{author}{Zhou, P.}, \bibinfo{author}{Yang,
  Z.}, \bibinfo{author}{Liu, X.}, \bibinfo{author}{Grundy, J.C.},
  \bibinfo{year}{2018}.
\newblock \bibinfo{title}{Fasttagrec: fast tag recommendation for software
  information sites}.
\newblock \bibinfo{journal}{Automated Software Engineering}
  \bibinfo{volume}{25}, \bibinfo{pages}{675--701}.
\bibitem[{Liu et~al.(2022)Liu, Yang, Chen and Yu}]{liu2022sotitle}
\bibinfo{author}{Liu, K.}, \bibinfo{author}{Yang, G.}, \bibinfo{author}{Chen,
  X.}, \bibinfo{author}{Yu, C.}, \bibinfo{year}{2022}.
\newblock \bibinfo{title}{Sotitle: A transformer-based post title generation
  approach for stack overflow}.
\newblock \bibinfo{journal}{arXiv preprint arXiv:2202.09789} .
\bibitem[{Loshchilov and Hutter(2017)}]{loshchilov2018decoupled}
\bibinfo{author}{Loshchilov, I.}, \bibinfo{author}{Hutter, F.},
  \bibinfo{year}{2017}.
\newblock \bibinfo{title}{Decoupled weight decay regularization}.
\newblock \bibinfo{journal}{arXiv preprint arXiv:1711.05101} .
\bibitem[{Lu et~al.(2021)Lu, Guo, Ren, Huang, Svyatkovskiy, Blanco, Clement,
  Drain, Jiang, Tang et~al.}]{Lu2021CodeXGLUEAM}
\bibinfo{author}{Lu, S.}, \bibinfo{author}{Guo, D.}, \bibinfo{author}{Ren, S.},
  \bibinfo{author}{Huang, J.}, \bibinfo{author}{Svyatkovskiy, A.},
  \bibinfo{author}{Blanco, A.}, \bibinfo{author}{Clement, C.},
  \bibinfo{author}{Drain, D.}, \bibinfo{author}{Jiang, D.},
  \bibinfo{author}{Tang, D.}, et~al., \bibinfo{year}{2021}.
\newblock \bibinfo{title}{Codexglue: A machine learning benchmark dataset for
  code understanding and generation}.
\newblock \bibinfo{journal}{arXiv preprint arXiv:2102.04664} .
\bibitem[{Mondal et~al.(2021)Mondal, Saifullah, Bhattacharjee, Rahman and
  Roy}]{mondal2021early}
\bibinfo{author}{Mondal, S.}, \bibinfo{author}{Saifullah, C.K.},
  \bibinfo{author}{Bhattacharjee, A.}, \bibinfo{author}{Rahman, M.M.},
  \bibinfo{author}{Roy, C.K.}, \bibinfo{year}{2021}.
\newblock \bibinfo{title}{Early detection and guidelines to improve unanswered
  questions on stack overflow}, in: \bibinfo{booktitle}{14th Innovations in
  Software Engineering Conference (formerly known as India Software Engineering
  Conference)}, pp. \bibinfo{pages}{1--11}.
\bibitem[{Papineni et~al.(2002)Papineni, Roukos, Ward and
  Zhu}]{Papineni2002BleuAM}
\bibinfo{author}{Papineni, K.}, \bibinfo{author}{Roukos, S.},
  \bibinfo{author}{Ward, T.}, \bibinfo{author}{Zhu, W.J.},
  \bibinfo{year}{2002}.
\newblock \bibinfo{title}{Bleu: a method for automatic evaluation of machine
  translation}, in: \bibinfo{booktitle}{Proceedings of the 40th annual meeting
  of the Association for Computational Linguistics}, pp.
  \bibinfo{pages}{311--318}.
\bibitem[{Radford et~al.()Radford, Narasimhan, Salimans and
  Sutskever}]{radford2018improving}
\bibinfo{author}{Radford, A.}, \bibinfo{author}{Narasimhan, K.},
  \bibinfo{author}{Salimans, T.}, \bibinfo{author}{Sutskever, I.}, .
\newblock \bibinfo{title}{Improving language understanding by generative
  pre-training} .
\bibitem[{Raffel et~al.(2020)Raffel, Shazeer, Roberts, Lee, Narang, Matena,
  Zhou, Li and Liu}]{raffel2019exploring}
\bibinfo{author}{Raffel, C.}, \bibinfo{author}{Shazeer, N.},
  \bibinfo{author}{Roberts, A.}, \bibinfo{author}{Lee, K.},
  \bibinfo{author}{Narang, S.}, \bibinfo{author}{Matena, M.},
  \bibinfo{author}{Zhou, Y.}, \bibinfo{author}{Li, W.}, \bibinfo{author}{Liu,
  P.J.}, \bibinfo{year}{2020}.
\newblock \bibinfo{title}{Exploring the limits of transfer learning with a
  unified text-to-text transformer}.
\newblock \bibinfo{journal}{Journal of Machine Learning Research}
  \bibinfo{volume}{21}, \bibinfo{pages}{1--67}.
\bibitem[{Robertson and Zaragoza(2009)}]{robertson2009probabilistic}
\bibinfo{author}{Robertson, S.}, \bibinfo{author}{Zaragoza, H.},
  \bibinfo{year}{2009}.
\newblock \bibinfo{title}{The probabilistic relevance framework: Bm25 and
  beyond}.
\newblock \bibinfo{journal}{Information Retrieval} \bibinfo{volume}{3},
  \bibinfo{pages}{333--389}.
\bibitem[{Rubei et~al.(2020)Rubei, Sipio, Nguyen, Rocco and
  Ruscio}]{Rubei2020PostFinderMS}
\bibinfo{author}{Rubei, R.}, \bibinfo{author}{Sipio, C.D.},
  \bibinfo{author}{Nguyen, P.T.}, \bibinfo{author}{Rocco, J.D.},
  \bibinfo{author}{Ruscio, D.D.}, \bibinfo{year}{2020}.
\newblock \bibinfo{title}{Postfinder: Mining stack overflow posts to support
  software developers}.
\newblock \bibinfo{journal}{Information and Software Technology}
  \bibinfo{volume}{127}, \bibinfo{pages}{106367}.
\bibitem[{See et~al.(2017)See, Liu and Manning}]{See2017GetTT}
\bibinfo{author}{See, A.}, \bibinfo{author}{Liu, P.J.},
  \bibinfo{author}{Manning, C.D.}, \bibinfo{year}{2017}.
\newblock \bibinfo{title}{Get to the point: Summarization with
  pointer-generator networks}, in: \bibinfo{booktitle}{Proceedings of the 55th
  Annual Meeting of the Association for Computational Linguistics (Volume 1:
  Long Papers)}, pp. \bibinfo{pages}{1073--1083}.
\bibitem[{Shi et~al.(2021)Shi, Wang, Du, Zhang, Han, Zhang and
  Sun}]{Shi2021CASTEC}
\bibinfo{author}{Shi, E.}, \bibinfo{author}{Wang, Y.}, \bibinfo{author}{Du,
  L.}, \bibinfo{author}{Zhang, H.}, \bibinfo{author}{Han, S.},
  \bibinfo{author}{Zhang, D.}, \bibinfo{author}{Sun, H.}, \bibinfo{year}{2021}.
\newblock \bibinfo{title}{Cast: Enhancing code summarization with hierarchical
  splitting and reconstruction of abstract syntax trees}.
\newblock \bibinfo{journal}{arXiv preprint arXiv:2108.12987} .
\bibitem[{Shi et~al.(2022)Shi, Fried, Ghazvininejad, Zettlemoyer and
  Wang}]{Shi2022NaturalLT}
\bibinfo{author}{Shi, F.}, \bibinfo{author}{Fried, D.},
  \bibinfo{author}{Ghazvininejad, M.}, \bibinfo{author}{Zettlemoyer, L.},
  \bibinfo{author}{Wang, S.I.}, \bibinfo{year}{2022}.
\newblock \bibinfo{title}{Natural language to code translation with execution}.
\newblock \bibinfo{journal}{arXiv preprint arXiv:2204.11454} .
\bibitem[{Tang et~al.(2021)Tang, Li, Ge, Shen, Zhu and Luo}]{tang2021ast}
\bibinfo{author}{Tang, Z.}, \bibinfo{author}{Li, C.}, \bibinfo{author}{Ge, J.},
  \bibinfo{author}{Shen, X.}, \bibinfo{author}{Zhu, Z.}, \bibinfo{author}{Luo,
  B.}, \bibinfo{year}{2021}.
\newblock \bibinfo{title}{Ast-transformer: Encoding abstract syntax trees
  efficiently for code summarization}, in: \bibinfo{booktitle}{2021 36th
  IEEE/ACM International Conference on Automated Software Engineering (ASE)},
  \bibinfo{organization}{IEEE Computer Society}. pp.
  \bibinfo{pages}{1193--1195}.
\bibitem[{Tang et~al.(2022)Tang, Shen, Li, Ge, Huang, Zhu and
  Luo}]{Tang2022ASTTransCS}
\bibinfo{author}{Tang, Z.}, \bibinfo{author}{Shen, X.}, \bibinfo{author}{Li,
  C.}, \bibinfo{author}{Ge, J.}, \bibinfo{author}{Huang, L.},
  \bibinfo{author}{Zhu, Z.}, \bibinfo{author}{Luo, B.}, \bibinfo{year}{2022}.
\newblock \bibinfo{title}{Ast-trans: Code summarization with efficient
  tree-structured attention}.
\newblock \bibinfo{journal}{2022 IEEE/ACM 44th International Conference on
  Software Engineering (ICSE)} , \bibinfo{pages}{150--162}.
\bibitem[{Tu et~al.(2016)Tu, Lu, Liu, Liu and Li}]{Tu2016ModelingCF}
\bibinfo{author}{Tu, Z.}, \bibinfo{author}{Lu, Z.}, \bibinfo{author}{Liu, Y.},
  \bibinfo{author}{Liu, X.}, \bibinfo{author}{Li, H.}, \bibinfo{year}{2016}.
\newblock \bibinfo{title}{Modeling coverage for neural machine translation}.
\newblock \bibinfo{journal}{arXiv preprint arXiv:1601.04811} .
\bibitem[{Vaswani et~al.(2017)Vaswani, Shazeer, Parmar, Uszkoreit, Jones,
  Gomez, Kaiser and Polosukhin}]{Vaswani2017AttentionIA}
\bibinfo{author}{Vaswani, A.}, \bibinfo{author}{Shazeer, N.},
  \bibinfo{author}{Parmar, N.}, \bibinfo{author}{Uszkoreit, J.},
  \bibinfo{author}{Jones, L.}, \bibinfo{author}{Gomez, A.N.},
  \bibinfo{author}{Kaiser, {\L}.}, \bibinfo{author}{Polosukhin, I.},
  \bibinfo{year}{2017}.
\newblock \bibinfo{title}{Attention is all you need}.
\newblock \bibinfo{journal}{Advances in neural information processing systems}
  \bibinfo{volume}{30}.
\bibitem[{Wang et~al.(2014)Wang, Lo, Vasilescu and
  Serebrenik}]{Wang2014EnTagRecAE}
\bibinfo{author}{Wang, S.}, \bibinfo{author}{Lo, D.},
  \bibinfo{author}{Vasilescu, B.}, \bibinfo{author}{Serebrenik, A.},
  \bibinfo{year}{2014}.
\newblock \bibinfo{title}{Entagrec++: An enhanced tag recommendation system for
  software information sites}.
\newblock \bibinfo{journal}{Empirical Software Engineering}
  \bibinfo{volume}{23}, \bibinfo{pages}{800--832}.
\bibitem[{Wang et~al.(2022)Wang, Wei, Schuurmans, Le, Chi and
  Zhou}]{wang2022self}
\bibinfo{author}{Wang, X.}, \bibinfo{author}{Wei, J.},
  \bibinfo{author}{Schuurmans, D.}, \bibinfo{author}{Le, Q.},
  \bibinfo{author}{Chi, E.}, \bibinfo{author}{Zhou, D.}, \bibinfo{year}{2022}.
\newblock \bibinfo{title}{Self-consistency improves chain of thought reasoning
  in language models}.
\newblock \bibinfo{journal}{arXiv preprint arXiv:2203.11171} .
\bibitem[{Wang et~al.(2015)Wang, Xia and Lo}]{Wang2015TagCombineRT}
\bibinfo{author}{Wang, X.}, \bibinfo{author}{Xia, X.}, \bibinfo{author}{Lo,
  D.}, \bibinfo{year}{2015}.
\newblock \bibinfo{title}{Tagcombine: Recommending tags to contents in software
  information sites}.
\newblock \bibinfo{journal}{Journal of Computer Science and Technology}
  \bibinfo{volume}{30}, \bibinfo{pages}{1017--1035}.
\bibitem[{Wang et~al.(2021)Wang, Wang, Joty and Hoi}]{Wang2021CodeT5IU}
\bibinfo{author}{Wang, Y.}, \bibinfo{author}{Wang, W.}, \bibinfo{author}{Joty,
  S.}, \bibinfo{author}{Hoi, S.C.}, \bibinfo{year}{2021}.
\newblock \bibinfo{title}{Codet5: Identifier-aware unified pre-trained
  encoder-decoder models for code understanding and generation}.
\newblock \bibinfo{journal}{arXiv preprint arXiv:2109.00859} .
\bibitem[{Xia et~al.(2013)Xia, Lo, Wang and Zhou}]{Xia2013TagRI}
\bibinfo{author}{Xia, X.}, \bibinfo{author}{Lo, D.}, \bibinfo{author}{Wang,
  X.}, \bibinfo{author}{Zhou, B.}, \bibinfo{year}{2013}.
\newblock \bibinfo{title}{Tag recommendation in software information sites}.
\newblock \bibinfo{journal}{2013 10th Working Conference on Mining Software
  Repositories (MSR)} , \bibinfo{pages}{287--296}.
\bibitem[{Xu et~al.(2021)Xu, Hoang, Sharma, Yang, Xia and
  Lo}]{Xu2021Post2VecLD}
\bibinfo{author}{Xu, B.}, \bibinfo{author}{Hoang, T.}, \bibinfo{author}{Sharma,
  A.}, \bibinfo{author}{Yang, C.}, \bibinfo{author}{Xia, X.},
  \bibinfo{author}{Lo, D.}, \bibinfo{year}{2021}.
\newblock \bibinfo{title}{Post2vec: Learning distributed representations of
  stack overflow posts}.
\newblock \bibinfo{journal}{IEEE Transactions on Software Engineering} ,
  \bibinfo{pages}{1--1}.
\bibitem[{Xu et~al.(2022)Xu, Alon, Neubig and Hellendoorn}]{Xu2022ASE}
\bibinfo{author}{Xu, F.F.}, \bibinfo{author}{Alon, U.},
  \bibinfo{author}{Neubig, G.}, \bibinfo{author}{Hellendoorn, V.J.},
  \bibinfo{year}{2022}.
\newblock \bibinfo{title}{A systematic evaluation of large language models of
  code}, in: \bibinfo{booktitle}{Proceedings of the 6th ACM SIGPLAN
  International Symposium on Machine Programming}, pp. \bibinfo{pages}{1--10}.
\bibitem[{Zhang et~al.(2022)Zhang, Keung, Yu, Xie, Yang, Ma and
  Zhang}]{zhang2022improving}
\bibinfo{author}{Zhang, F.}, \bibinfo{author}{Keung, J.}, \bibinfo{author}{Yu,
  X.}, \bibinfo{author}{Xie, Z.}, \bibinfo{author}{Yang, Z.},
  \bibinfo{author}{Ma, C.}, \bibinfo{author}{Zhang, Z.}, \bibinfo{year}{2022}.
\newblock \bibinfo{title}{Improving stack overflow question title generation
  with copying enhanced codebert model and bi-modal information}.
\newblock \bibinfo{journal}{Information and Software Technology} ,
  \bibinfo{pages}{106922}.
\bibitem[{Zhou et~al.(2019)Zhou, Liu, Liu, Yang and Grundy}]{zhou2019deep}
\bibinfo{author}{Zhou, P.}, \bibinfo{author}{Liu, J.}, \bibinfo{author}{Liu,
  X.}, \bibinfo{author}{Yang, Z.}, \bibinfo{author}{Grundy, J.},
  \bibinfo{year}{2019}.
\newblock \bibinfo{title}{Is deep learning better than traditional approaches
  in tag recommendation for software information sites?}
\newblock \bibinfo{journal}{Information and software technology}
  \bibinfo{volume}{109}, \bibinfo{pages}{1--13}.
\bibitem[{Zhou et~al.(2017)Zhou, Liu, Yang and Zhou}]{Zhou2017ScalableTR}
\bibinfo{author}{Zhou, P.}, \bibinfo{author}{Liu, J.}, \bibinfo{author}{Yang,
  Z.}, \bibinfo{author}{Zhou, G.}, \bibinfo{year}{2017}.
\newblock \bibinfo{title}{Scalable tag recommendation for software information
  sites}.
\newblock \bibinfo{journal}{2017 IEEE 24th International Conference on Software
  Analysis, Evolution and Reengineering (SANER)} , \bibinfo{pages}{272--282}.
\bibitem[{Zhou et~al.(2022a)Zhou, Shen, Zhang, Yang, Han and
  Chen}]{Zhou2022AutomaticSC}
\bibinfo{author}{Zhou, Y.}, \bibinfo{author}{Shen, J.}, \bibinfo{author}{Zhang,
  X.}, \bibinfo{author}{Yang, W.}, \bibinfo{author}{Han, T.},
  \bibinfo{author}{Chen, T.}, \bibinfo{year}{2022}a.
\newblock \bibinfo{title}{Automatic source code summarization with graph
  attention networks}.
\newblock \bibinfo{journal}{Journal of Systems and Software}
  \bibinfo{volume}{188}, \bibinfo{pages}{111257}.
\bibitem[{Zhou et~al.(2022b)Zhou, Yu, Fan, Huang and
  Yang}]{Zhou2022SummarizingSC}
\bibinfo{author}{Zhou, Z.}, \bibinfo{author}{Yu, H.}, \bibinfo{author}{Fan,
  G.}, \bibinfo{author}{Huang, Z.}, \bibinfo{author}{Yang, X.},
  \bibinfo{year}{2022}b.
\newblock \bibinfo{title}{Summarizing source code with hierarchical code
  representation}.
\newblock \bibinfo{journal}{Information and Software Technology}
  \bibinfo{volume}{143}, \bibinfo{pages}{106761}.

\end{thebibliography}

\end{document}